\documentclass[12pt,manuscript,psfig,epsfig,pdffig,superscriptaddress,floatfix,nofootinbib]{revtex4}

\usepackage[latin9]{inputenc}
\usepackage{amsmath}
\usepackage{amssymb}
\usepackage{graphicx}
\usepackage{epstopdf}
\usepackage{pstricks}
\usepackage{color}

\makeatletter
\@ifundefined{textcolor}{}
{%
	\definecolor{BLACK}{gray}{0}
	\definecolor{WHITE}{gray}{1}
	\definecolor{RED}{rgb}{1,0,0}
	\definecolor{GREEN}{rgb}{0,1,0}
	\definecolor{BLUE}{rgb}{0,0,1}
	\definecolor{CYAN}{cmyk}{1,0,0,0}
	\definecolor{MAGENTA}{cmyk}{0,1,0,0}
	\definecolor{YELLOW}{cmyk}{0,0,1,0}
}

\usepackage{dcolumn}
\usepackage{bm}
\usepackage{amsfonts}
\usepackage{cancel}

\makeatother

\begin{document}

\title{Unambiguous cancellation of divergences of   $H \to \gamma \gamma$ process via one W loop in unitary gauge: \\  Gauge invariance in Dyson scheme and
 physical boundary condition}







\author{Shi-Yuan Li}
\affiliation{School of Physics, Shandong University, Jinan 250100, P. R. China}
\author{Zong-Guo Si}
\affiliation{School of Physics, Shandong University, Jinan 250100, P. R. China}


\author{Xiao-Feng Zhang}  
 \affiliation{School of Physics, Shandong University, Jinan 250100, P. R. China}

\begin{abstract}
Following the thread of R. Gastmans, S. L. Wu and T. T. Wu, the calculation in the unitary gauge   for the $H \to \gamma \gamma$ process
via one W loop
is repeated, but
 without the specific choice of  the  independent  loop momentum for  the  Feynman diagrams.  This is based on the original 'Dyson scheme' provided in Dyson's classical paper.
 I.e., the original integrations on all  propagator momenta are kept,  not expressed
 by the specified independent loop momentum.  Correspondingly,
  the 4-dimension $\delta$ function at each vertex  in which      
 the 4-momentum conservation  is embedded, is retained. 
 Together with  the Ward identity of the W-W-photon vertex, the 4-momentum conservation of each vertex guarantee the cancellation of all
   divergent integrals  worse than logarithmic without   any uncertainty or ambiguity, with  any shift of integrated momentum eschewed. The calculation is in 4-dimension Minkowski phase space and without any help of regularization. The resulting integrals are to the most logarithmically divergent, hence is invariant
for  various setting of the independent loop momentum and any of its shift.  At last the  logarithmically  divergent symmetric (tensor) integration is determined by   the boundary condition at infinity in momentum space inherent of   the free
Feynman propagator, and the gauge (both $SU(2)\times U_Y(1)$ and $U_{em}(1)$) invariant  finite result can be obtained without the introduction of the 'Dyson subtraction'.
The physical boundary conditions at infinity of phase space can make sense in many problems.


\end{abstract}


\date{\today}
\maketitle	
\section{Introduction}\label{sec1}
The Glashow-Weinberg-Salam 
    electroweak (EW) theory is a SU(2)$\times$U(1) Yang-Mills gauge theory, with the gauge symmetry 'broken' by a scalar field via the Englert-Brout-Higgs-Guralnik-Hagen-Kibble Mechanism. This has been confirmed
from experiments, i.e., the massive $W^\pm$, $Z$ particles versus the massless photon, and a 'remaining' neutral scalar particle which is generally referred to as the Higgs particle, all have been well measured.  The EW theory is remarkably different from the other set of the
standard model, quantum chromodynamics (QCD) with its SU(3) gauge symmetry not broken. Though the confinement mechanism is yet
not fully understood, itself is a manifestation   of the non-broken---simply that all physical states are colour singlet, i.e., invariant w.r.t. the SU(3) group;
and any  physical observable constructed with the 'colourful' quark gluon degree of freedom must be SU(3) invariant.

In general, a realistic calculation of the S-matrix or scattering amplitude employing the quantized  field theory
of the standard model need to fix a specific gauge and it is adopted that the result  should be  independent from the choice of the gauge.
For the feasibility of loop calculations, the 'renormalizable gauge', e.g., $R_{\xi}$ gauge, is favoured. Both EW and QCD can take this gauge. On the other hand,  for the EW theory, one can also take a specific 'physical gauge',  unitary gauge,
 where only the physical degrees of freedom present. According to Weinberg \cite{wein73}, this gauge can be defined by imposing the condition of the scalar field relevant to its vacuum expectation value and the 'symmetry breaking'. Such a gauge may not be introduced for QCD.  
   This does not matter as long as that all physical observables like decay width, cross section, etc., are   the same for any gauges. However, recently,  
   possible  implication for the 'contrary'  seems recognized, based on the careful revisit on the $H \to \gamma \gamma$ decay width in the unitary gauge and $R_{\xi}$ gauge \cite{Wu:2017rxt,Wu:2016nqf,Gastmans:2015vyh,Gastmans:2011ks,Gastmans:2011wh}.
   The first indication of this fact is that calculations in the unitary gauge for loop diagrams should be extensively studied to eliminate any uncertainty.


  The main purpose of this paper is to  %
 repeat the calculation in unitary gauge
by Gastmans,  Wu and Wu in \cite{Gastmans:2011ks,Gastmans:2011wh} on the $H \to \gamma \gamma$ process (in this paper we refer to these two papers and the works in them as GWW)
to get some insights on the possibility of  elimination of the uncertainties of this process as well as to find subtle implication for likely  problems, i.e., divergence cancellation to get reasonable and reliable physical prediction.   
 In the unitary gauge, high   divergence order terms  of loop diagrams appear and  should properly cancel. Or else one can not get the correct  result, either not possible to make the comparison with results from other gauges.
For any diagram whose divergence worse than logarithmic, to shift the integrated momentum can lead to extra terms with lower divergency (or finite).
In such case,   the complete set of diagrams at certain order with correct inter-relations of the  loop momenta must be treated together
 to get the correct result, as pointed by GWW. Only a   part    of the set of  diagrams shifting the momenta will change the result.
  However, 
we can start from the 'original expression' of each of the   Feynman diagrams  introduced in  Dyson's  classical paper \cite{Dyson:1949ha} that all the momenta
  of the propagators  are kept as  the original,   not  to be  expressed by the  independent  loop momentum such as that of  GWW in a correlated way. These momenta are  related by the 4-momentum  conservation of each vertex, expressed by the $\delta$ functions respectively attached with each vertex.  Hence  the integrations on all these momenta are also  kept.  By investigating  the details of the divergence cancellation we confirm
  the inter-relation of the  loop momenta of these diagrams (hence the symmetry relations of the diagrams) pointed by GWW as a  natural result.  In these  high order ($> log$) divergences the cancellation is of no uncertainty. 
 This is the  standard perturbative expansion on the S-matrix, there is no ambiguity of the assignment of the loop momenta from  the starting point. And any  shift of integrated  momentum in  integrals of divergence worse than logarithmic  is eschewed (the logarithmic  surface term and the corresponding  boundary condition of Feynman propagator specially  investigated in the following).
 Redefinition (with new uncertainty introduced) from the original theoretical framework by Dyson \cite{Dyson:1949ha} is also not necessary in this process.
 At last the  $SU(2) \times U(1)$ gauge invariant finite result is naturally obtained without the Dyson subtraction used by of GWW.
                %
    %
Therefore,  regularization is also not necessary for the calculations, which are all done in 4-dimension Minkowski phase space.
As the final result is finite,  without the introduction of
renormalization,  hence  no extra (physical) renormalization condition is introduced.
In this kind of finite case the unambiguity of the Dyson scheme is quite nontrivial  to give definite prediction  
{\it from divergence cancellation without  renormalization. }
  Such a study from this most general/original footing, i.e., the  most general way of setting   the integrated momenta,   has yet not available in literature  until now. This method can be definitely applied to $H \to \gamma Z $ process
via  W loop. It can  shed light on cases of  more complex diagrams and/or with more loops  to find  the
  proper   inter-relation of the  loop momenta of those diagrams. 





 In more details,  the calculation on Higgs particle decay into two photons via the W loop in the unitary gauge encounters two kinds of divergent terms in each of the diagrams:

  (1) Those worse than logarithmic, which need the exact cancellation  and only logarithmic
divergent (and/or finite) terms can remain. In our procedure,
 their cancellation is the result that different diagrams give exactly the same  integral(s)  with opposite sign. So the cancellation is on the  integral level rather than the integrand level. This is the key point for these worse divergences. All these need not the setting of the specific
integrated momentum. The special inter-relation, e.g.,   that of GWW,  is the natural result of the general relation and the cancellation without the need of being set \`{a} priori. 
In our derivation,  shift of integrated variables is carefully eschewed.

 (2) The other divergence is the logarithmic terms \footnote{Here we would like to emphasize that these terms are not {\it superficially by power counting},  but frankly, logarithmic. The case here for triangle diagrams is the most simple one. For more complex cases, the divergence can be much more complex, especially worse than simply power counting (a famous example is the photon-photon scattering one loop box diagram), so that more labour is needed to separate the divergences worse than logarithmic and to obtain the really logarithmic one.},  and the choice of the loop momenta can be independent for each diagram since shift of the momenta will
 not change the result now. Indeed, the most easy calculation way is to employ  the usual Feynman-Schwinger parametrization
 to deal with the proper part of the terms and to get the result.
The  cancellation of this divergence need the symmetric (tensor) integration. This is the key point for  this case, the boundary condition  need to be carefully considered to determine the surface term which is not definite \cite{Ferreira:2011cv} for various boundary condition. We argue here,
the one  inherent in 
the  Dyson scheme  (without any redefinition of the propagator by some regularization) can lead to the unambiguous result with the guarantee of the gauge invariance (obtaining the result same as that of the $R_\xi$ gauge).
The Dyson subtraction is not needed.

  %


In this paper we eschew  setting  independent loop momentum, as well as eschew any shift of variable for high divergence order integral by the help of the   early and original covariant purterbative quantum field theory (QFT)  S matrix theory, the  Dyson scheme. What is more we hence also 'eschew' the debate of different observation on gauge non-invariance (our result IS gauge INvariant) or uncertainties for this specific problem.
 For discussion on the uncertainty of QFT, see \cite{Duch:2020was} for a review and comments on debates of the specific problem. But in the Dyson scheme  all are definite and further redefinition on QFT or employing  regularization  (often introducing extra uncertainty \cite{Duch:2020was} ) is shown not necessary for this specific problem; besides, the uncertainty of surface term  \cite{Ferreira:2011cv} can be determined by the inherent boundary condition when consistently employing the Dyson scheme, as mentioned above.

    The content of this paper is as following:
 In Sec. 2,  
 divergences worse than  logarithmic  
  are all arranged and cancelled with the proper employment of the Dyson scheme, which also  confirm the GWW calculation with their inner relation of the diagrams and the independent loop momentum unambiguously  without the help of any regularization.  
 %
 %
 Then
 In Sec. 3, we will discuss the remaining logarithmic divergences, which will cancel to a finite result. 
We will give the result similar to the others,  by the analysis of the boundary condition of the propagator in 4 dimension Minkowski phase space. We will point out this  is also crucial to get other important results such as  the consistent result for the axial vector current conservation (anomaly free) \cite{Bao:2021byx}, as well calculation on  the H to gamma Z process \cite{LW}, etc.
and this implies deep physics related with the structure of space time. The results in unitary gauge also indicate that many investigation with loop diagrams could be done in unitary gauge for the advantage less diagrams and clear physical picture.
Some further information can be found in the appendices.

\section{Calculation}

{\large 1.}
  To say that a worse (than logartithmic) divergent  integral   is changed when shifting the integrated momentum, first of all one should ask, what is 'the original'  TO BE changed?  There may not exist the 'original' one for a single diagram, once taking into account that different Feynman diagrams are related and hence the loop momenta (\`{a} la GWW); however, one can have some definiteness starting  from the original form derived from the expansion of the S-matrix according  to the standard  Dyson-Wick procedure. Since the S-matrix is the exponential of a space-time integration; once taking these space-time integrations at   each order, one gets  $\delta$ functions, one  for each vertex, relating all the momenta of the propagators \cite{Dyson:1949ha} with the 4-momentum conservation for each vertex. If we start  from such a form for each diagram, without integrating the $\delta$ functions, there will be no indefiniteness. 

\begin{figure}[htb]
   	\centering
   	\begin{tabular}{cccccc}
   		\scalebox{0.3}[0.3]{\includegraphics{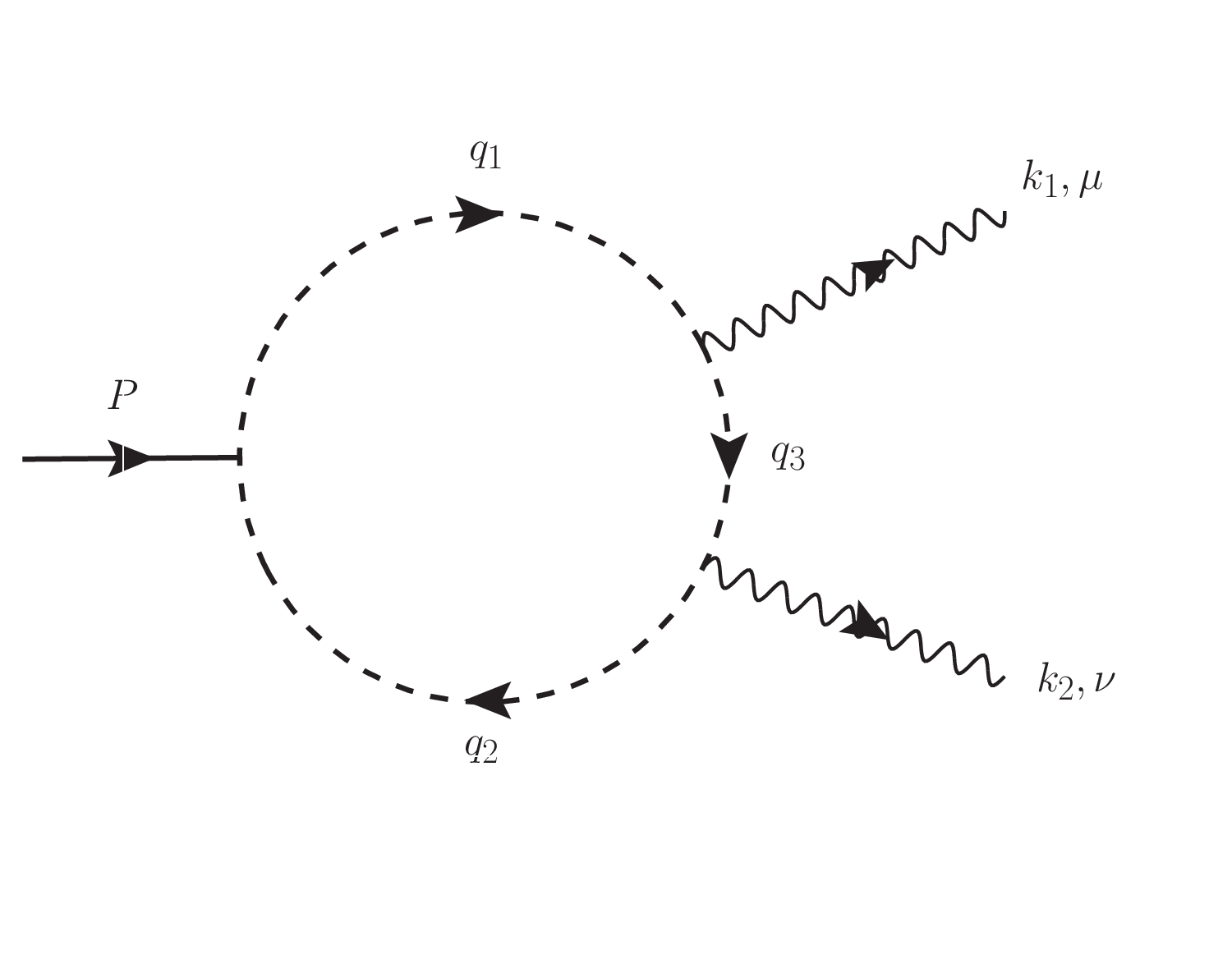}}&
   		\scalebox{0.3}[0.3]{\includegraphics{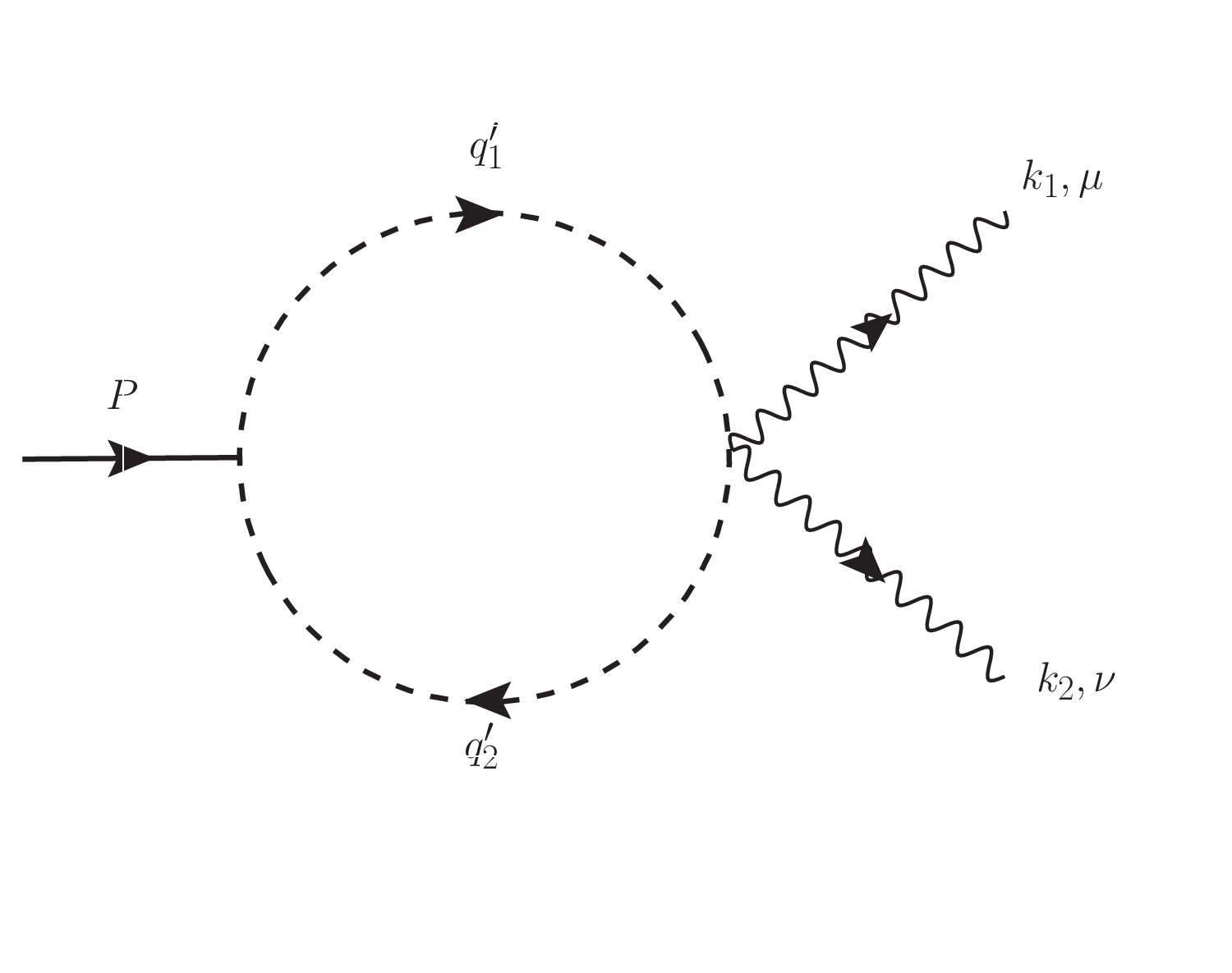}}&\\
   		{\scriptsize ($T_1$)}&{\scriptsize ($T_2$)}\\
   		\scalebox{0.33}[0.33]{\includegraphics{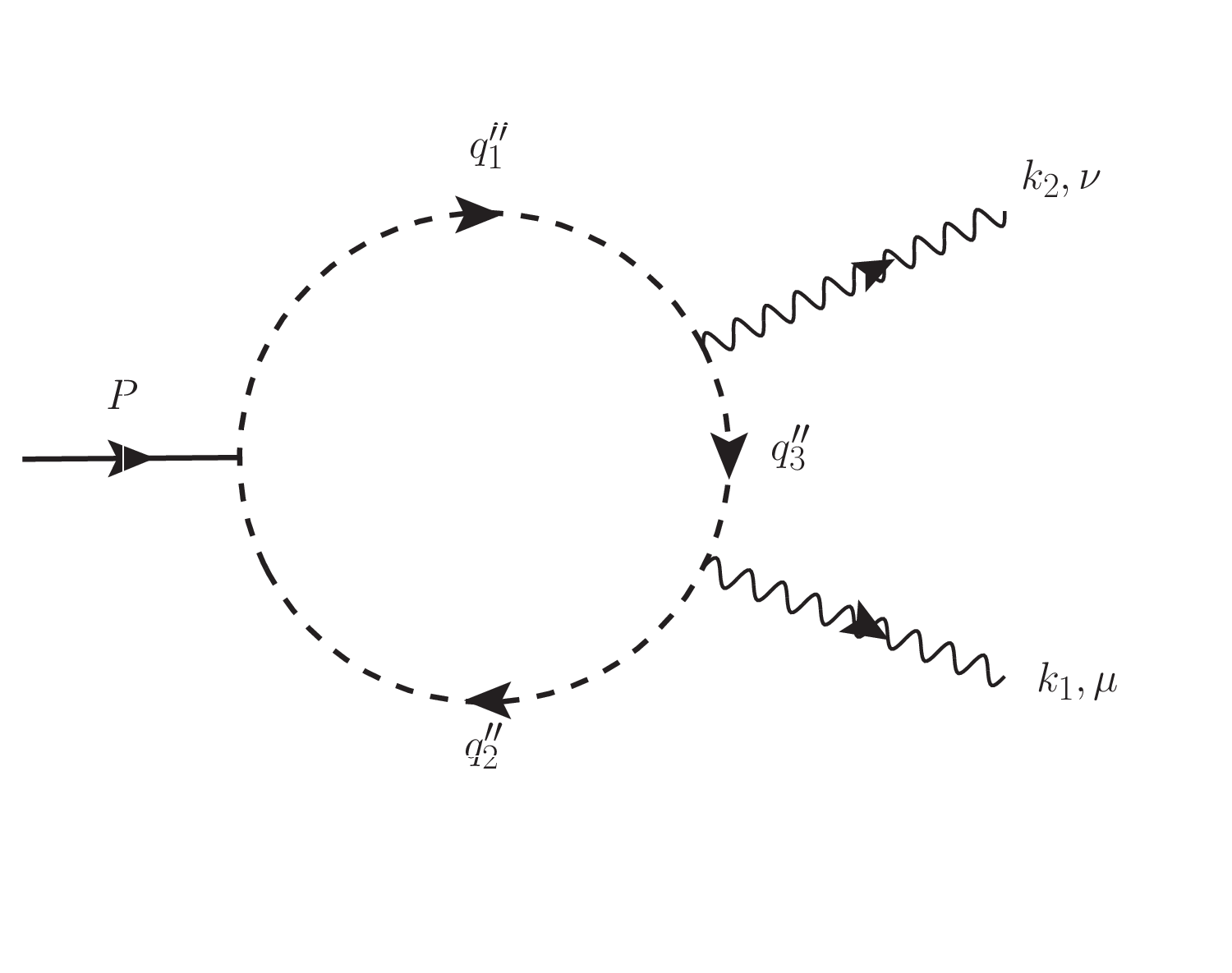}}&~\\
   		{\scriptsize ($T_3$)}& ~\\   		
   	\end{tabular}
   	\caption{The one W loop Feynman diagrams. 
   In general, the inner integrated momenta should be considered as not correlated between different diagrams, so here we mark those of $T_2$ and $T_3$ with prime or double primes. In the manuscript, this is also implied though not explicitly written. But for the purpose of cancellation which is determined by the whole  integral, they are taken to be the same in the concrete step of derivation. This is easily to be tracked.}\label{fd}
   \end{figure}

 As convention, The S-matrix and T-matrix have the relation $S=I+i T$, and the matrix element between initial and final states
 $i T_{fi}=i (2\pi)^4 \delta(P_f-P_i) \mathfrak{M}_{fi}$,
for the case without the presence of  an external classical field which  breaks the space-time displacement invariance.
Here  we retain all the  momenta respectively corresponding to  each propagator and hence all
$\delta$ functions respectively corresponding to each   vertex.  The one corresponding to  the initial-final state energy momentum conservation
is contained in these $\delta$ functions. After integrating over them, one will get the above form of T-matrix element and then the $\mathfrak{M}_{fi}$
is the integration of the independent loop momenta only,  without the $\delta$ functions attached to the vertices.
This is the standard procedure in developing the Dyson-Wick perturbation theory in the interaction picture \cite{Dyson:1949ha}.  The four-momentum conservation
$\delta$ function attached to each of the vertices is the result of integration  of space-time contained in the perturbative expansion of the
S-matrix, and is the manifestation of space-time displacement invariance \footnote{I.e., shifting space-time position not changing the result. So not only the boundary of space time at infinity matters,
but also the singularity, topology (e.g., hole), connection of different part of manifold (e.g., bubble wall) matter}.
In the following, we do the calculation and do not 'integrate out' the $\delta$ functions of each vertex until have to.
So here we deal with the matrix elements $T_{fi} $ rather than $\mathfrak{M}_{fi}$:
\begin{eqnarray}
\label{t1}
 T_1 &= &  \frac{-ie^2gM}{(2 \pi)^4} \int d^4q_1 d^4q_2 d^4q_3 (2 \pi)^4 \delta(P-q_1+q_2) \delta(q_1-k_1-q_3)\delta(q_3-k_2-q_2)   \\
~& \times & (g_\alpha~^{\beta} - \frac{q_{1\alpha} q_1 ^{\beta}}{M^2} ) (g ^{\rho \sigma}-\frac{q_3 ^\rho q_3 ^\sigma}{M^2} ) (g^{\alpha \gamma} -\frac{q_2^\alpha q_2 ^\gamma}{M^2} )
 \frac{V_{\beta \mu \rho} (q_1,  -k_1, -q_3) ~ V_{\sigma \nu \gamma}(q_3, -k_2, -q_2)} {(q_1 ^2-M^2)(q_3 ^2-M^2)(q_2 ^2-M^2)} .\nonumber
\end{eqnarray}
In the above Equation and following, symbols like $V_{\sigma \nu \gamma}(q_3, -k_2, -q_2)$ represents the $W\gamma W$ vertex just similar as GWW.  Here we do not explicitly write the matrix element subscript$~_{fi}$, and all the $\delta$ fuctions are understood as four-dimensional one,
i.e., $\delta(P_1-P_2) :=\delta^4 (P_1-P_2)$. As GWW, we also omit the photon polarization vectors, and  $T_1$ should be understood as $T_{1\mu \nu}$.
These integrations are easy to be considered  in 4 as well as in D dimension. In D dimension,  it is $d^4 q_1 d^4 q_2 d^4 q_3 (\delta^4 )^3$ $\rightarrow$ $d^D q_1 d^D q_2 d^D q_3 (\delta^D)^3$. Though we will not discuss the debates on dimensional regularization as mentioned above,  
  we just   clarify the procedure of cancellation in the whole section  also needed and valid in D dimension  
  to eliminate any ambiguity caused by the setting of loop momenta for the application of dimensional regularization.
\begin{eqnarray}
\label{t2}
 T_2 &= &  \frac{ie^2gM}{(2 \pi)^4} \int d^4q_1 d^4q_2  (2 \pi)^4 \delta(P-q_1+q_2) \delta(q_1-q_2-k_1-k_2)   \\
~& \times & (g_\alpha~^{\beta} - \frac{q_{1\alpha} q_1 ^{\beta}}{M^2} )  (g^{\alpha \gamma} -\frac{q_2^\alpha q_2 ^\gamma}{M^2} )
 \frac{2g_{\mu \nu} g_{\beta \gamma}-g_{\mu \beta} g_{\nu \gamma} -g_{\mu \gamma} g_{\nu \beta}} {(q_1 ^2-M^2)(q_2 ^2-M^2)} .\nonumber
\end{eqnarray}
\begin{eqnarray}
\label{t3}
 T_3 &= &  \frac{-ie^2gM}{(2 \pi)^4} \int d^4q_1 d^4q_2 d^4q_3 (2 \pi)^4 \delta(P-q_1+q_2) \delta(q_1-k_2-q_3)\delta(q_3-k_1-q_2)   \\
~& \times & (g_\alpha~^{\beta} - \frac{q_{1\alpha} q_1 ^{\beta}}{M^2} ) (g ^{\rho \sigma}-\frac{q_3 ^\rho q_3 ^\sigma}{M^2} ) (g^{\alpha \gamma} -\frac{q_2^\alpha q_2 ^\gamma}{M^2} )
 \frac{V_{\beta \nu \rho} (q_1,  -k_2, -q_3) ~ V_{\sigma \mu \gamma}(q_3, -k_1, -q_2)} {(q_1 ^2-M^2)(q_3 ^2-M^2)(q_2 ^2-M^2)} .\nonumber
\end{eqnarray}
These just correspond to    $M_{1,2,3}$ of GWW times the $ \delta$ function of the whole energy-momentum conservation. The momentum of each propagator
recover back to the  original one of the definition of the propagator. If integrating the $\delta$ functions, by choosing the loop  momenta as GWW do, one can get the similar
equations as Eqs. (2.2-2.4) of GWW \footnote{e.g., in $T_1$,  $q_1=k+(k_1+k_2)/2$, $q_2=k-(k_1+k_2)/2$, and $q_3  =k+(-k_1+k_2)/2$, respectively.}.
If we set the independent integrated momentum $k$ in other ways, we can get other forms.
Here   
  the relation between $T_1$ and $T_3$, i.e., $\mu \leftrightarrow \nu$, and $k_1 \leftrightarrow k_2$
is  clear to be read out.
    The     Feynman diagrams are shown in Figure \ref{fd}.
  This corresponds to that the momentum space Feynman rules  (see GWW) are slightly modified (in fact 'recovered', see the classical paper of Dyson \cite{Dyson:1949ha}, especially its Eq. (20) and discussions before and after it) as:  Any propagator with momentum $q$ has an extra $[\int \frac{d^4 q}{(2\pi)^4}]$ 'operator',  i.e., should this integration on $q$ to be done in the calculation of the Feynman diagram;
  any vertex has an extra factor $ (2\pi)^4 \delta(\sum_i q_i)$, with  $q_i$,  each momentum of all the propagators meeting in the vertex,    incoming.

In this way, and at this beginning step,  one has no requirement  of choosing the independent loop momentum (momenta for multi-loop case) for
each diagram 
 \footnote{If one variable only appear in the $\delta$ functions for one special term, it is allowed to be integrated out.}.
This is EXACTLY the definition of the amplitude without any ambiguity, provided that one adopts each term in the Dyson-Wick  perturbation  theory is well defined and the
propagators and vertices are well defined. 
 And this is just the Feynman rules itself.
Not to keep  the integration on each propagator momentum and the $\delta$ function of each vertex is the 'simplified version' of Feynman rules, by integrating  out the $\delta$ functions, only retaining the one for the initial-final state four momentum conservation.  
  The independent momenta has been set inexplicitly by the insight on the relations of all the diagrams corresponding to a process, such as those in GWW.
Some old quantum field theory text books in fact gave the original/not-simplified version, as is used here. The simplified one is only popular in modern time.
The key point for the  latter is just that one has to set the independent integration momenta $\grave{a}~ priori$.  This is just  what we want to elude.  So we start from the 'old' version.
To our knowledge, no textbook (or even Dyson's paper \cite{Dyson:1949ha}) persuaded to keep the $\delta$ functions until they have to be integrated  as we will do in the following,  this is one
of the reasons why we write  in details in this section.

In the Dyson-Wick perturbative  theory framework  and in the interaction picture, the propagator is  definite (provided properly defined in relation with the standard scattering theory, and we will discuss in Sec. 3), the $\delta$ functions from each vertex has set the relations of all the propagator
momenta definitely. This means once one keeps all these relations and  calculate according to mathematics without mistakes, one has no uncertainty (or one is free) from from choosing  
independent 'loop'  momenta for each diagram or some of the terms.

This also means that if at some step to write down the amplitude with further 'freedom/restriction' which is not included in the above Feynman rules,
 especially the propagator integrations and vertex $\delta$ functions, or is contradicting to this definite result,
 this step may introduce more things deviating from the exact
definition of the amplitude in the above standard  framework. So that step  is a wrong step rather than some 'ambiguity',
  since it is not included in the theory in the  beginning. 


{\large 2.} To better illustrate the calculation procedure, here we mention  the  formulae we use repeatedly in the calculation.
I.e., Eqs. (2.5-2.12) of GWW \cite{Gastmans:2011wh}. For these equations, we will refer to as GWW2.5, GWW2.6,..., GWW2.12 in this paper. Complementary to GWW2.9,
we  add an equation as $V_{\alpha \mu \gamma } (p_1, -k_1, p_3) p_3 ^\gamma=-(p_1^2 g_{\alpha \mu}-p_{1\alpha}p_{2\mu})$, i.e.,
the case contracted with the other inner momentum $p_3$.  In GWW, they only list the formula for
$ p_1 ^\alpha V_{\alpha \mu \gamma } (p_1, -k_1, p_3)$.
However, for simplicity, we still refer this complemented formulae as GWW2.9. 
Here we would like to address that all these formulations still valid in D dimension.
%
  The property of the  3-vertex, which is named as Ward Identity (W.I.) in GWW, is the key role in the calculation.
 For the feasibility to investigate the $H \to \gamma Z$ process   later 
 we  in fact list the equations for $H \to \gamma Z$ process
   corresponding to GWW (2.5-2.12)
  in Appendix A.  They straightforwardly  go to GWW (2.5-2.12) by simply taking $M_Z=0$.

We also would like to mention that, the explicitly employment of $k_{1 \mu} \epsilon^\mu= k_{2 \nu} \epsilon^\nu=0$  (GWW2.5) everywhere
means that it is valid without  any account of any divergent factors timing with it.

In 4 as well as in D dimension,
 one can
arrange each of the diagrams in terms associate with the minus power of the W mass  M (with the extra overall M factors from the coupling taken out, as GWW).
 Same as   the thread of GWW, the divergences are investigated according to the the inverse power of the W mass.
 We will also check  
 once they can cancel in 4 dimension,  whether  they can  cancel in D dimension (the terms can be taken as  convergent). This will be explicitly mentioned in the following.

First is
the $M^{-6}$ term, which only appears in $T_1$ and $T_3$, are explicitly read out  as  zero, respectively (without the need of cancellation), because of the property of the W.I., GWW2.11, 2.12.
So, in four dimension, this means that the divergence of power 6th does not exist, and even does not leave any terms with lower divergence which may draw any
uncertainty  to the following analysis.  This W.I., frankly the basic property of the three-boson vertex applied to the specific process we investigate,
can not be broken or made to any deviation.  This also is the case for D dimension,
 i.e.,  in D dimension,  
   terms proportional to $M^{-6}$ also equals to zero.
  All the following analysis respect this fact. Any direct result from the W.I.
without referring to any other relations is solid without any ambiguity.

{\large 3.}   $M^{-4}$ terms

In all the following, the $\frac{-ie^2gM}{(2 \pi)^4}$ factor will not explicitly written, and  all terms should multiply with this factor to get the
proper terms in the corresponding T amplitude of Eqs. (\ref{t1}-\ref{t3}).
So the $M^{-4}$ terms from Eq.{\ref{t1}} is (analogy to   Eq. (3.1) of GWW,  only with  the extra $\delta$ functions, integrations,  and without the overall constant factors mentioned  above)
\begin{eqnarray}
\label{t11}
T_{11}&=&\frac{1}{M^4} \int d^4q_1 d^4q_2 d^4q_3 (2 \pi)^4 \delta(P-q_1+q_2) \delta(q_1-k_1-q_3)\delta(q_3-k_2-q_2)\\
~ & \times &  q_{1\alpha} q_1 ^{\beta} g^{\rho \sigma} q_2 ^{\alpha} q_2 ^{\gamma} \frac{V_{\beta \mu \rho} (q_1,  -k_1, -q_3) ~ V_{\sigma \nu \gamma}(q_3, -k_2, -q_2)} {(q_1 ^2-M^2)(q_3 ^2-M^2)(q_2 ^2-M^2)}. \nonumber
\end{eqnarray}
Now since $q_1 ^{\beta} V_{\beta \mu \rho} (q_1,  -k_1, -q_3)= (q_3^2-M^2) g_{\mu \rho}-q_{3\mu}q_{3\rho}+M^2 g_{\mu \rho}$, according to GWW2.10,
 $T_{11}=T_{111}+T_{112}+T_{113}$.

In this form,  $T_{112}$
\begin{eqnarray}
\label{t112}
T_{112}&=&\frac{1}{M^4} \int d^4q_1 d^4q_2 d^4q_3 (2 \pi)^4 \delta(P-q_1+q_2) \delta(q_1-k_1-q_3)\delta(q_3-k_2-q_2)\\
~ & \times &  q_{1\alpha}  q_2 ^{\alpha} q_2 ^{\gamma} (-q_{3\mu}q_{3\rho})g^{\rho \sigma}\frac{V_{\sigma \nu \gamma}(q_3, -k_2, -q_2)} {(q_1 ^2-M^2)(q_3 ^2-M^2)(q_2 ^2-M^2)}, \nonumber
\end{eqnarray}
is explicitly read to be  zero, since the factor $-q_2 ^{\gamma} q_{3\rho} g^{\rho \sigma} V_{\sigma \nu \gamma}(q_3, -k_2, -q_2)=0 $ according to GWW2.12.
 \begin{eqnarray}
 \label{t111}
T_{111} &=&\frac{1}{M^4} \int d^4q_1 d^4q_2 d^4q_3 (2 \pi)^4 \delta(P-q_1+q_2) \delta(q_1-k_1-q_3)\delta(q_3-k_2-q_2)\\
~ & \times &  q_{1 \alpha}  q_2 ^{\alpha} q_2 ^{\gamma} g_{\mu \rho} g^{\rho \sigma} \frac{V_{\sigma \nu \gamma}(q_3, -k_2, -q_2)} {(q_1 ^2-M^2)(q_2 ^2-M^2)}. \nonumber
\end{eqnarray}
Again to employ GWW2.9 (our compensary) for  $V_{\sigma \nu \gamma}(q_3, -k_2, -q_2) q_2 ^{\gamma}$,
it is
\begin{eqnarray}
T_{111} &=&\frac{1}{M^4} \int d^4q_1 d^4q_2 d^4q_3 (2 \pi)^4 \delta(P-q_1+q_2) \delta(q_1-k_1-q_3)\delta(q_3-k_2-q_2)\\
~ & \times &   \frac{ q_1 \cdot  q_2  (q_3^2 g_{\mu \nu}-q_{3\mu}q_{3\nu} )} {(q_1 ^2-M^2)(q_2 ^2-M^2)}. \nonumber 
\end{eqnarray}
In this step, one can take into account that, $\int dx f(x) \delta(x-a)=\int dx f(a) \delta(x-a)$, to use the relation $q_3=q_1-k_1=q_2+k_2$,
to get $$ (q_3^2 g_{\mu \nu}-q_{3\mu}q_{3\nu} )= q_1 \cdot q_2 g_{\mu \nu} -q_{1\mu} q_{2\nu} +(q_1\cdot k_2-q_2\cdot k_1-k_1\cdot k_2) g_{\mu\nu}.$$
In fact, all the Ward identities for the 3-boson vertex we use here also have employed the energy momentum conservation at the vertex.  So
\begin{eqnarray}
T_{111} &=&\frac{1}{M^4} \int d^4q_1 d^4q_2 d^4q_3 (2 \pi)^4 \delta(P-q_1+q_2) \delta(q_1-k_1-q_3)\delta(q_3-k_2-q_2)\\
~ & \times &    q_1 \cdot  q_2 ~~ \frac{q_1 \cdot q_2 g_{\mu \nu} -q_{1\mu} q_{2\nu} +(q_1\cdot k_2-q_2\cdot k_1-k_1\cdot k_2) g_{\mu\nu}} {(q_1 ^2-M^2)(q_2 ^2-M^2)}. \nonumber 
\end{eqnarray}
Now we check the $M^{-4}$ term in  $T_3$, which is called $T_{31}$, and find the similar derivation can be done, to get $T_{31}=T_{311}+T_{312}+T_{313}$.
 $T_{312}=0$,  according to GWW2.12, and
\begin{eqnarray}
\label{t311}
T_{311} &=&\frac{1}{M^4} \int d^4q_1 d^4q_2 d^4q_3 (2 \pi)^4 \delta(P-q_1+q_2) \delta(q_1-k_2-q_3)\delta(q_3-k_1-q_2)\\
~ & \times &    q_1 \cdot  q_2 ~~ \frac{q_1 \cdot q_2 g_{\mu \nu} -q_{2\mu} q_{1\nu} +(q_1\cdot k_1-q_2\cdot k_2-k_1\cdot k_2) g_{\mu\nu}} {(q_1 ^2-M^2)(q_2 ^2-M^2)}. \nonumber 
\end{eqnarray}
It is easy to see that in $T_{111}$ and $T_{311}$, $q_3$ only appears in the $\delta$ functions and can be integrated to get 1, and  this integration leads no any restriction for the remaining integrand and momenta. Now the delta function is the same for $T_{111}$ and $T_{311}$,
 so to take the  momenta of the corresponding propagators in $T_1$ and $T_2$ as the same is the result of the expression, e.g., that of the
 vertex and its contraction with the  special 'momentum over mass'   term of unitary gauge,  hence the
 symmetric relation between these 2 diagrams. This  is
 explicitly expressed here and has been employed before calculation as criteria for setting the independent integrated momentum in GWW.
 Here we see that two contractions respectively with the two vertices   both lead to terms of $q_3$ and can be expressed symmetrically by $q_1$ and $q_2$.
 All the above we never add any requirement, restriction or redefinition/seeting on the momenta, all the relations are those inherent in the integrand (including the $\delta$ functions).

 By definition, for any finite $q_1$,  $q_2$ and   $q_3$,
  the corresponding $\delta$ function can be always integrated to get 1.
And since the trivial contribution of infinite values of the propagator momenta (see discussions in the following on symmetrical integration),
there is no  singularity to
introduce extra term by this definition.    
  Consequently 
  the integrands in $T_{111}$ and $T_{311}$ are not dependent on  $q_3$.
So   after integrating $q_3$ and a $\delta$ function 
  these two integrals  are well defined and can be summed together,  and then sum with $M^{-4}$ terms from $T_2$, with the corresponding integrated  propagator momenta taken as the same (since we combine the whole well-defined  integral not just integrand only, and the cancellation here is at the integral level rather than integrand level like GWW).
The final result is definitely  zero  without uncertainty:
 \begin{eqnarray}
T_{111} +T_{311}&=&\frac{1}{M^4} \int d^4q_1 d^4q_2  (2 \pi)^4 \delta(P-q_1+q_2) \delta(q_1-q_2-k_1-k_2)\\
~ & \times &    q_1 \cdot  q_2 ~~ \frac{2 q_1 \cdot q_2 g_{\mu \nu} -q_{1\mu} q_{2\nu}-q_{2\mu} q_{1\nu} } {(q_1 ^2-M^2)(q_2 ^2-M^2)},  \nonumber 
\end{eqnarray}
since $ (q_1\cdot k_2-q_2\cdot k_1-k_1\cdot k_2) g_{\mu\nu}+ (q_1\cdot k_1-q_2\cdot k_2-k_1\cdot k_2) g_{\mu\nu}=0$, takeing now $q_1-q_2=k_1+k_2$ from the $\delta$ function.
 The $M^{-4}$ term in $T_2$ is 
 \begin{eqnarray}
T_{21}&=&\frac{-1}{M^4} \int d^4q_1 d^4q_2  (2 \pi)^4 \delta(P-q_1+q_2) \delta(q_1-q_2-k_1-k_2)\\
~ & \times &    q_1 \cdot  q_2 ~~ \frac{2 q_1 \cdot q_2 g_{\mu \nu} -q_{1\mu} q_{2\nu}-q_{2\mu} q_{1\nu} } {(q_1 ^2-M^2)(q_2 ^2-M^2)},  \nonumber 
\end{eqnarray}
so $$T_{111} +T_{311}+T_{21}=0.$$
The 'formal' calculation in fact is more simple than the explicit one with the independent loop momentum as in GWW.


As mentioned above, the expression of integrand of  $T_{111}$ and $T_{311}$ are exactly expressed by $q_1, q_2$,
 and the   $\delta$ functions are exactly the same form when integrating over  $q_3$.  Then the summation is  just negative to $T_{21}$, as the {\it whole} integral (the integrand, as well as  the integrated variables constrained by the same $\delta$ functions).
 So the cancellation is definite, just read from the    expressions themselves.
 This general result---result from the most general expression without setting the momenta beforehand---'enforces' the set of propagator momenta by GWW when using the independent integrated momentum (or other equivalent ways). If some one deliberately set one of the above three terms from the three diagrams with  a different  momentum ($q_1$ or $q_2$) from the other two, and since these are  high divergent terms, this may lead $T_{111} +T_{311}+T_{21}$ to be low divergent  or finite non-zero value.
   %
   In D($<$ 4, small enough) dimension, these are considered as NOT divergent terms, and the shift seems  allowed, and  each
 of the terms itself can be calculated individually. However, extra pole of D or finite terms at D to 4 limit may appear. 
 So the above relation and cancellation must be taken first  even for D dimension.
  %
  %
 Here we would like to draw the attention of the reader that since all the metric tensor appears in this part with an
external index (i.e., $\mu$  or $\nu$) which to be contracted with the photon polarization vectors,  one will not get any terms dependent on the dimension, D,
from the self-contraction of the metric tensor. So the calculations for the integrands  are similar for D as well as 4 dimensions.
 This argument is also valid for $M^{-2}$ terms but need to be checked for $M^{0}$ terms  for there would appear self contraction of the metric tensor,  which equals to D rather than 4 in D dimension.

Now it is seen that the 4th power divergence exactly cancels, and no finite (nonzero) term proportional to $M^{-4}$.
The remaining terms from $T_{11}$ and $T_{31}$ is $T_{113}$ and $T_{313}$ which are proportional to $M^{-2}$:
 \begin{eqnarray}
 \label{t113}
T_{113} &=&\frac{1}{M^2} \int d^4q_1 d^4q_2 d^4q_3 (2 \pi)^4 \delta(P-q_1+q_2) \delta(q_1-k_1-q_3)\delta(q_3-k_2-q_2)\\
~ & \times &   \frac{ q_1 \cdot  q_2 V_{\mu \nu \gamma}(q_3, -k_2, -q_2) q_2^{\gamma}} {(q_1 ^2-M^2)(q_3 ^2-M^2)(q_2 ^2-M^2)}. \nonumber
\end{eqnarray}
 \begin{eqnarray}
T_{313} &=&\frac{1}{M^2} \int d^4q_1 d^4q_2 d^4q_3 (2 \pi)^4 \delta(P-q_1+q_2) \delta(q_1-k_2-q_3)\delta(q_3-k_1-q_2)\\
~ & \times &   \frac{ q_1 \cdot  q_2 V_{\nu \mu \gamma}(q_3, -k_1, -q_2) q_2^{\gamma}} {(q_1 ^2-M^2)(q_3 ^2-M^2)(q_2 ^2-M^2)}. \nonumber  \label{t313}
\end{eqnarray}
They are to be considered together with other $M^{-2}$ terms.

{\large 4.}   $M^{-2}$ terms

Besides the above  $M^{-2}$ terms, we need to investigate the following:  From $T_1$
\begin{eqnarray}
\label{t12}
T_{12} &= &  \frac{-1}{M^2} \int d^4q_1 d^4q_2 d^4q_3 (2 \pi)^4 \delta(P-q_1+q_2) \delta(q_1-k_1-q_3)\delta(q_3-k_2-q_2)   \\
~& \times & g_\alpha~^{\beta}  g ^{\rho \sigma} q_2^\alpha q_2 ^\gamma
 \frac{V_{\beta \mu \rho} (q_1,  -k_1, -q_3) ~ V_{\sigma \nu \gamma}(q_3, -k_2, -q_2)} {(q_1 ^2-M^2)(q_3 ^2-M^2)(q_2 ^2-M^2)}, \nonumber
\end{eqnarray}
\begin{eqnarray}
\label{t13}
 T_{13} &= &  \frac{-1}{M^2} \int d^4q_1 d^4q_2 d^4q_3 (2 \pi)^4 \delta(P-q_1+q_2) \delta(q_1-k_1-q_3)\delta(q_3-k_2-q_2)   \\
~& \times & g_\alpha~^{\beta}  q_3 ^\rho q_3 ^\sigma g^{\alpha \gamma}
 \frac{V_{\beta \mu \rho} (q_1,  -k_1, -q_3) ~ V_{\sigma \nu \gamma}(q_3, -k_2, -q_2)} {(q_1 ^2-M^2)(q_3 ^2-M^2)(q_2 ^2-M^2)}, \nonumber
\end{eqnarray}
\begin{eqnarray}
\label{t14}
 T_{14} &= &  \frac{-1}{M^2} \int d^4q_1 d^4q_2 d^4q_3 (2 \pi)^4 \delta(P-q_1+q_2) \delta(q_1-k_1-q_3)\delta(q_3-k_2-q_2)   \\
~& \times & q_{1\alpha} q_1 ^{\beta}g ^{\rho \sigma}g^{\alpha \gamma}
 \frac{V_{\beta \mu \rho} (q_1,  -k_1, -q_3) ~ V_{\sigma \nu \gamma}(q_3, -k_2, -q_2)} {(q_1 ^2-M^2)(q_3 ^2-M^2)(q_2 ^2-M^2)}. \nonumber
\end{eqnarray}

The term from $T_2$ is
\begin{eqnarray}
\label{t22e3}
 T_{22+23} &= &  \frac{1}{M^2} \int d^4q_1 d^4q_2  (2 \pi)^4 \delta(P-q_1+q_2) \delta(q_1-q_2-k_1-k_2)   \\
~& \times &  \frac{2q_1^2 g_{\mu \nu} + 2q_2^2 g_{\mu \nu}-2 q_{1 \mu} q_{1 \nu}-2 q_{2 \mu} q_{2 \nu}} {(q_1 ^2-M^2)(q_2 ^2-M^2)} \nonumber
\end{eqnarray}
corresponding to (M22+M23 of GWW).

We see $T_{12}$ and $T_{14}$ both give $(q_3^2-M^2)$ factor in numerator when  the Ward identity directly applied, which  can reduce the corresponding factor in
denominator and combine with $T_{22+23}$.
The following showes the fact.

As above, by applying WI GWW2.10 to  $V_{\sigma \nu \gamma}(q_3, -k_2, -q_2) q_2 ^\gamma$ and $q_1 ^{\beta} V_{\beta \mu \rho} (q_1,  -k_1, -q_3)$ respectively to  $T_{12}$ and $T_{14}$,  both are written with 3 terms, respectively:
$T_{12}=T_{121}+T_{122}+T_{123}$ and $T_{14}=T_{141}+T_{142}+T_{143}$.

This discussion also applies to $T_3$, so the following is to investigate $$T_{121}+T_{141}+T_{321}+T_{341}+T_{22+23}.$$
\begin{eqnarray}
\label{t121}
T_{121} &= &  \frac{-1}{M^2} \int d^4q_1 d^4q_2 d^4q_3 (2 \pi)^4 \delta(P-q_1+q_2) \delta(q_1-k_1-q_3)\delta(q_3-k_2-q_2)   \\
~& \times &  q_2^\beta
 \frac{V_{\beta \mu \nu} (q_1,  -k_1, -q_3)} {(q_1 ^2-M^2)(q_2 ^2-M^2)}. \nonumber
\end{eqnarray}
For this part, we use
$q_2=q_1-P$,  so can use W.I. again, while the extra term  $-P^\beta
 V_{\beta \mu \nu} (q_1,  -k_1, -q_3)$  will combine with the corresponding extra term from $T_{141}$, since
 \begin{eqnarray}
 \label{t141}
 T_{141} &= &  \frac{-1}{M^2} \int d^4q_1 d^4q_2 d^4q_3 (2 \pi)^4 \delta(P-q_1+q_2) \delta(q_1-k_1-q_3)\delta(q_3-k_2-q_2)   \\
~& \times &
 \frac{V_{\mu \nu \gamma} (q_3,  -k_2, -q_2)} {(q_1 ^2-M^2)(q_2 ^2-M^2)}q_1^\gamma. \nonumber
\end{eqnarray}
We use $q_1=q_2+P$, and the extra term is $V_{\mu \nu \gamma} (q_3,  -k_2, -q_2)P^\gamma $.
\begin{eqnarray}
T_{121}+T_{141} &= &  \frac{-1}{M^2} \int d^4q_1 d^4q_2 d^4q_3 (2 \pi)^4 \delta(P-q_1+q_2) \delta(q_1-k_1-q_3)\delta(q_3-k_2-q_2)   \\
~& \times &
 \frac{2q_3^2 g_{\mu \nu} - 2 q_{3 \mu}q_{3\nu}+2 k_1 \cdot k_2 g_{\mu \nu}+3 q_{3 \mu}k_{1 \nu}-3 k_{2\mu}q_{3 \nu}-4k_{2 \mu}k_{1\nu}} {(q_1 ^2-M^2)(q_2 ^2-M^2)} \nonumber \\
~ &= &  \frac{-1}{M^2} \int d^4q_1 d^4q_2 (2 \pi)^4 \delta(P-q_1+q_2) \delta(q_1-q_2-k_1-k_2) \nonumber  \\
~& ~&  \hspace{-2.6cm}\times
 \frac{(q_1^2+q_2^2) g_{\mu \nu} -2 k_1 \cdot q_1 g_{\mu \nu}+2 k_2 \cdot q_2 g_{\mu \nu}+2 k_1 \cdot k_2 g_{\mu \nu}- 2 q_{1 \mu}q_{2\nu}+3 q_{1 \mu}k_{1 \nu}-3 k_{2\mu}q_{2 \nu}-4k_{2 \mu}k_{1\nu}} {(q_1 ^2-M^2)(q_2 ^2-M^2)} \nonumber 
\end{eqnarray}
Here $q_{3\mu}=q_{1\mu}$, and $q_{3\nu}=q_{2\nu}$. Two $  q_3^2$~'s are expressed as functions of $q_1^2$ and $q_2^2$ respectively to get a more  symmetric form.
Then we integrate over $q_3$ to get 1 for the second step.
\begin{eqnarray}
T_{321}+T_{341}
~ &= &  \frac{-1}{M^2} \int d^4q_1 d^4q_2 (2 \pi)^4 \delta(P-q_1+q_2) \delta(q_1-q_2-k_1-k_2)   \\
~& ~&  \hspace{-2.6cm}\times
 \frac{(q_1^2+q_2^2) g_{\mu \nu} -2 k_2 \cdot q_1 g_{\mu \nu}+2 k_1 \cdot q_2 g_{\mu \nu}+2 k_1 \cdot k_2 g_{\mu \nu}- 2 q_{1 \nu}q_{2\mu}+3 q_{1 \nu}k_{2 \mu}-3 k_{1\nu}q_{2 \mu}-4k_{2 \mu}k_{1\nu}} {(q_1 ^2-M^2)(q_2 ^2-M^2)} \nonumber 
\end{eqnarray}
From the above two equations and  Eq.(\ref{t22e3}), one can easily found
$$T_{121}+T_{141}+T_{321}+T_{341}+T_{22+23}=0,$$
by observing $2 q_{1 \mu} q_{1 \nu}+2 q_{2 \mu} q_{2 \nu}- 2 q_{1 \mu}q_{2\nu}- 2 q_{1 \nu}q_{2\mu}=2k_{2 \mu}k_{1\nu}$.
This is  also valid for D dimension.

~~~~~~


~~~~~~~~~~~~~~

~~~~~~~~~~~~~~~~~~~~~~~~~

Now the remaining $M^{-2}$ terms are all from $T_1$ and $T_3$.  
Those from $T_{1}$ are  
 $$T_{113}+T_{13}+T_{122}+T_{142}.$$
$T_{113}$ and $T_{13}$ are shown in Eq.(\ref{t113}) and Eq.(\ref{t13}), respectively. They can be directly calculated with the W.I. of GWW2.9, say, applying to
$V_{\mu \nu \gamma}(q_3, -k_2, -q_2) q_2^{\gamma}$,
 $V_{\beta \mu \rho} (q_1,  -k_1, -q_3) q_3 ^\rho, $ and  $q_3 ^\sigma V_{\sigma \nu \gamma}(q_3, -k_2, -q_2)$.
For Eqs.(\ref{t12}, \ref{t14}), we have discussed their first term $T_{121}$ and $T_{141}$, while the second terms  of them are
\begin{eqnarray}
T_{122} &= &  \frac{1}{M^2} \int d^4q_1 d^4q_2 d^4q_3 (2 \pi)^4 \delta(P-q_1+q_2) \delta(q_1-k_1-q_3)\delta(q_3-k_2-q_2)   \\
~& \times &  q_2^\beta
 \frac{V_{\beta \mu \sigma} (q_1,  -k_1, -q_3) q_3 ^\sigma q_{3\nu}} {(q_1 ^2-M^2)(q_3 ^2-M^2)(q_2 ^2-M^2)}, \nonumber 
\end{eqnarray}
\begin{eqnarray}
T_{142} &= &  \frac{1}{M^2} \int d^4q_1 d^4q_2 d^4q_3 (2 \pi)^4 \delta(P-q_1+q_2) \delta(q_1-k_1-q_3)\delta(q_3-k_2-q_2)   \\
~& \times &
 \frac{q_{3\mu} q_3 ^\sigma V_{\sigma \nu \gamma} (q_3,  -k_2, -q_2) } {(q_1 ^2-M^2)(q_3 ^2-M^2)(q_2 ^2-M^2)} q_1^\gamma. \nonumber 
\end{eqnarray}
We here again  apply the W.I. and combine them together to obtain a simple form
\begin{eqnarray}
~ & ~ & T_{113}+T_{13}+T_{122}+T_{142} \\
~& = & \frac{1}{M^2} \int d^4q_1 d^4q_2 d^4q_3 \frac{  (2 \pi)^4 \delta(P-q_1+q_2) \delta(q_1-k_1-q_3)\delta(q_3-k_2-q_2) }  {(q_1 ^2-M^2)(q_3 ^2-M^2)(q_2 ^2-M^2)} \nonumber \\
 ~& \times & (2 q_1^2 q_{2\mu} q_{2\nu} + 2 q_2^2 q_{1\mu} q_{1\nu} -4 q_1\cdot q_2 q_{1\mu} q_{2\nu} + q_1\cdot q_2 q_3^2 g_{\mu \nu}-q_1^2 q_2 ^2 g_{\mu \nu}). \nonumber
\end{eqnarray}
It looks as quadratic, but easy to see  in fact to the most linear,
since

$2 q_1\cdot q_2=-(q_1-q_2)^2+q_1^2+q_2^2 $,  then
$ (2 q_1^2 q_{2\mu} q_{2\nu} + 2 q_2^2 q_{1\mu} q_{1\nu} -4 q_1\cdot q_2 q_{1\mu} q_{2\nu})$   equals to
 $$2 q_1^2(q_{2\mu}-q_{1\mu})q_{2\nu} +2 q_2^2 q_{1\mu}(q_{1\nu}-q_{2\nu})
+2 (q_1-q_2)^2 q_{1\mu}q_{2 \nu}=2 q_1^2 (-k_{2\mu})q_{2\nu} +2 q_2^2 q_{1\mu}k_{1\nu}
+2 (k_1+k_2)^2 q_{1\mu}q_{2 \nu};$$

and
$q_1\cdot q_2 q_3^2 g_{\mu \nu}-q_1^2 q_2 ^2 g_{\mu \nu}=-\frac{(k_1+k_2)^2}{2}q_3^2 g_{\mu \nu}+\frac{q_1^2+q_2^2}{2}q_3^3 g_{\mu \nu}-q_1^2 q_2 ^2 g_{\mu \nu}.$

However, $\frac{q_1^2+q_2^2}{2}q_3^3 g_{\mu \nu}-q_1^2 q_2 ^2 g_{\mu \nu}= (q_1^2 q_2\cdot k_2 -q_2^2 q_1\cdot k_1)g_{\mu \nu}$ hence is also linear.


Now we write
$T_{113}+T_{13}+T_{122}+T_{142}$  
 as the summation of two parts:
 \begin{eqnarray}
 \label{line}
  &  & \frac{1}{M^2} \int d^4q_1 d^4q_2 d^4q_3 \frac{  (2 \pi)^4 \delta(P-q_1+q_2) \delta(q_1-k_1-q_3)\delta(q_3-k_2-q_2) }  {(q_1 ^2-M^2)(q_3 ^2-M^2)(q_2 ^2-M^2)} \nonumber \\
 ~& \times & 2 q_1^2 (-k_{2\mu})q_{2\nu} +2 q_2^2 q_{1\mu}k_{1\nu}+ (q_1^2 q_2\cdot k_2 -q_2^2 q_1\cdot k_1)g_{\mu \nu}, 
\end{eqnarray}
\begin{eqnarray}
\label{loog}
  &  & \frac{1}{M^2} \int d^4q_1 d^4q_2 d^4q_3 \frac{  (2 \pi)^4 \delta(P-q_1+q_2) \delta(q_1-k_1-q_3)\delta(q_3-k_2-q_2) }  {(q_1 ^2-M^2)(q_3 ^2-M^2)(q_2 ^2-M^2)} \nonumber \\
 ~& \times & 2 (k_1+k_2)^2 q_{1\mu}q_{2 \nu}-\frac{(k_1+k_2)^2}{2}q_3^2 g_{\mu \nu}, 
\end{eqnarray}
i.e., the linear and logarithmic divergent terms respectively.
Then the above linear term, after further taking out   logarithmic  and  finite terms from it,
 should  combine with the corresponding term from $T_3$, then is deduced   to get  as two terms,  one is logarithmic divergent, the other is finite.

 Some details are:
 \begin{eqnarray}
  2 q_1^2 (-k_{2\mu})q_{2\nu}&=&-2(q_3^2-M^2)k_{2\mu}q_{2\nu}-4q_3\cdot k_1 k_{2\mu}q_{2\nu}-2M^2k_{2\mu}q_{2\nu},  \nonumber \\
 2 q_2^2 q_{1\mu}k_{1\nu}&=& 2 (q_3^2-M^2) q_{1\mu}k_{1\nu} -4 q_3\cdot k_2 q_{1\mu}k_{1\nu} +2 M^2  q_{1\mu}k_{1\nu}, \nonumber\\
 (q_1^2 q_2\cdot k_2 -q_2^2 q_1\cdot k_1)g_{\mu \nu}&=&(q_3^2-M^2)(q_2\cdot k_2 -q_1 \cdot k_1) g_{\mu \nu}
 + 4 q_2\cdot k_2  q_1 \cdot k_1 g_{\mu \nu} + M^2(q_2\cdot k_2 -q_1 \cdot k_1) g_{\mu \nu}.  \nonumber
  \end{eqnarray}

So it is the  following linear term (with $(q_3^2-M^2)$ factor reduced with the common one in the denominator, and $q_3$ integrated  )
\begin{equation}
   \frac{1}{M^2} \int d^4q_1 d^4q_2  \frac{  (2 \pi)^4 \delta(P-q_1+q_2) \delta(q_1-q_2-k_1-k_2)}  {(q_1 ^2-M^2)(q_2 ^2-M^2)}
 ( -2k_{2\mu}q_{2\nu} +2  q_{1\mu}k_{1\nu} + (q_2\cdot k_2 -q_1 \cdot k_1) g_{\mu \nu})
 \end{equation}
 to be combined with that from $T_3$:
\begin{equation}
   \frac{1}{M^2} \int d^4q_1 d^4q_2  \frac{  (2 \pi)^4 \delta(P-q_1+q_2) \delta(q_1-q_2-k_1-k_2)}  {(q_1 ^2-M^2)(q_2 ^2-M^2)}
 ( -2k_{1\nu}q_{2\mu} +2  q_{1\nu}k_{2\mu} + (q_2\cdot k_1 -q_1 \cdot k_2) g_{\mu \nu}),
 \end{equation}
and deduces to logarithmic term. {\it Half} of their summation is then:
\begin{equation}
  \frac{1}{M^2} \int d^4q_1 d^4q_2  \frac{  (2 \pi)^4 \delta(P-q_1+q_2) \delta(q_1-q_2-k_1-k_2)}  {(q_1 ^2-M^2)(q_2 ^2-M^2)}
( 2  k_{2\mu}k_{1\nu} - k_1\cdot k_2  g_{\mu \nu}).
\end{equation}
This term can again be separated into a logarithmic term and a finite term,
\begin{eqnarray}
 &~& \frac{1}{M^2} \int d^4q_1 d^4q_2 d^4q_3 \frac{  (2 \pi)^4 \delta(P-q_1+q_2) \delta(q_1-k_1-q_3)\delta(q_3-k_2-q_2) }  {(q_1 ^2-M^2)(q_3 ^2-M^2)(q_2 ^2-M^2)}
q_3^2 ( 2  k_{2\mu}k_{1\nu} - k_1\cdot k_2  g_{\mu \nu}) \nonumber \\
&+&\frac{1}{M^2} \int d^4q_1 d^4q_2 d^4q_3 \frac{  (2 \pi)^4 \delta(P-q_1+q_2) \delta(q_1-k_1-q_3)\delta(q_3-k_2-q_2) }  {(q_1 ^2-M^2)(q_3 ^2-M^2)(q_2 ^2-M^2)}
(-M^2) ( 2  k_{2\mu}k_{1\nu} - k_1\cdot k_2  g_{\mu \nu}). \nonumber
 \end{eqnarray}
 Hence {\it effectively}
 $T_{113}+T_{13}+T_{122}+T_{142} =T1LG+T1F$, with
 \begin{eqnarray}
 \label{t1lg}
 T1LG = \frac{1}{M^2} \int d^4q_1 d^4q_2 d^4q_3 \frac{  (2 \pi)^4 \delta(P-q_1+q_2) \delta(q_1-k_1-q_3)\delta(q_3-k_2-q_2) }  {(q_1 ^2-M^2)(q_3 ^2-M^2)(q_2 ^2-M^2)}~~~~~~&~\\
 \times  [(-2 q_3^2 k_1\cdot k_2 + 4 k_1 \cdot q_3 k_2 \cdot q_3) g_{\mu \nu}-4 k_1 \cdot q_3 k_{2 \mu} q_{3 \nu}-4 k_2 \cdot q_3 q_{3 \mu} k_{1 \nu}
 +2 q_3^2 k_{2\mu} k_{1\nu} +4 k_1 \cdot k_2 q_{3\mu} q_{3 \nu}];&~ \nonumber
 \end{eqnarray}
and
  \begin{eqnarray}
 T1F&=&  \int d^4q_1 d^4q_2 d^4q_3 \frac{  (2 \pi)^4 \delta(P-q_1+q_2) \delta(q_1-k_1-q_3)\delta(q_3-k_2-q_2) }  {(q_1 ^2-M^2)(q_3 ^2-M^2)(q_2 ^2-M^2)}\\
 &\times& [2 q_{1\mu} k_{1\nu}-2k_{2\mu}q_{2\nu}-2k_{2\mu}k_{1\nu} +(q_2\cdot k_2-q_1 \cdot k_1+k_1\cdot k_2  )g_{\mu \nu}]. \nonumber
 \end{eqnarray}
The summation in fact is deduced into a quite simple form.

 It can be checked, by the product with the the overall factor,  the finite integral T1F is exactly GWW3.42 ($M_{1132}$) when  taking the choice  of integrated
loop momentum as GWW. The logarithmic one T1LG is  also easy to recover GWW3.41, when  taking the choice  of integrated
loop momentum as GWW. It is in fact that  employing  whichever independent loop momentum, by the help of Feynman-Schwinger parametrization,
one can always get the similar result.   Zero as GWW if only deal in 4 dimension and with a periodic boundary condition without taking into account the surface term determined by the physical boundary condition (see next section), or the usual finite result
 employing dimensional regularization \cite{SVVZ12, HTW, MZW,Jegerlehner:2011jm, SZC, CCNS, pit} as others. The discussion and reason why and how one gets the non zero result as dimensional regularization, in four dimension Minkowski spacetime and a 'natural' boundary condition of Dyson Scheme,  will be given in  next section.


 The calculation and result are  the same in D dimension. 
We would like to remind that the 4 appear in the $M^{-2}$ terms 
 is the result of combination of 4 terms, not from the dimension.
In the logarithmic $M^{0}$ terms, things could be different.

Here we also see that the quadratic divergent terms are deduced directly to logarithmic, by summing $T_1$ and $T_3$. 
 The linear divergent terms
 are all safely combined and summed to get logarithmic ones,  according to the momentum relations, without the requirement of the average on k $\leftrightarrow$ -k
 as in GWW.


   From the derivation of the cancellation of the linear divergence, we see it is quite non trivial as to separate and recombine the terms. As the above, it again  shows the subtle loop momentum relation exist between $T_1$ and $T_3$, which is
correctly set before calculation by GWW, and then after the average on k and -k, the linear divergence cancelled while the remaining logarithmic term are correct.

If we do not care about this, directly deal with the  $T_{113}+T_{13}+T_{122}+T_{142}$, Eq. (24) or as the summation of linear
and logarithmic terms (\ref{line}) + (\ref{loog}) in D dimension, where D smaller than 4 and the integral is finite.
Since it is finite, we directly employ Feynman-Schwinger parametrization and do the shift.
We find that the linear divergence is of no problem cancelled.  Terms giving logarithmic divergence for D $\to $ 4
and proportional to $1/M^2$ appear. But  it is not T1LG which can give FINITE result when  D $\to $ 4; it is just divergence not
able to be cancelled (by those from $T_3$).
This again  means that even one considers the extrapolation to D dimension, the relation of loop momenta still need to be respected hence divergent terms worse than logarithmic cancelled. The dimension regularization can  not fully take the  place of that. 

{\large 5.} $M^0$ terms

The third part of $T_{12},~ T_{14}$, i.e., $T_{123}, ~ T_{143}$, as well as the corresponding ones of $T_3$, are $M^0$ terms and to be investigated here
together with the corresponding terms from $T_2$  and the other $M^0$ terms from $T_1$ and $T_3$  
  (Pay attention that $T_2$ lack of a overall minus sign):
\begin{eqnarray}
T_{123} &= &  - \int d^4q_1 d^4q_2 d^4q_3 (2 \pi)^4 \delta(P-q_1+q_2) \delta(q_1-k_1-q_3)\delta(q_3-k_2-q_2)   \\
~& \times &  q_2^\beta
 \frac{V_{\beta \mu \sigma} (q_1,  -k_1, -q_3) g^\sigma~ \hspace{-0.26cm} _\nu} {(q_1 ^2-M^2)(q_3 ^2-M^2)(q_2 ^2-M^2)} \nonumber 
\end{eqnarray}
\begin{eqnarray}
T_{143} &= &  - \int d^4q_1 d^4q_2 d^4q_3 (2 \pi)^4 \delta(P-q_1+q_2) \delta(q_1-k_1-q_3)\delta(q_3-k_2-q_2)   \\
~& \times &
 \frac{g_\mu ~\hspace{-0.2cm}^\sigma V_{\sigma \nu \gamma} (q_3,  -k_2, -q_2) } {(q_1 ^2-M^2)(q_3 ^2-M^2)(q_2 ^2-M^2)} q_1^\gamma \nonumber 
\end{eqnarray}
\begin{eqnarray}
T_{15}&=&\int d^4q_1 d^4q_2 d^4q_3 (2 \pi)^4 \delta(P-q_1+q_2) \delta(q_1-k_1-q_3)\delta(q_3-k_2-q_2)  \\
~&\times & g_\alpha~^{\beta} g ^{\rho \sigma} g^{\alpha \gamma}
 \frac{V_{\beta \mu \rho} (q_1,  -k_1, -q_3) ~ V_{\sigma \nu \gamma}(q_3, -k_2, -q_2)} {(q_1 ^2-M^2)(q_3 ^2-M^2)(q_2 ^2-M^2)}\nonumber
\end{eqnarray}
\begin{eqnarray}
 T_{24} &= & - \int d^4q_1 d^4q_2  (2 \pi)^4 \delta(P-q_1+q_2) \delta(q_1-q_2-k_1-k_2)   \\
~& \times & g_\alpha~^{\beta}  g^{\alpha \gamma}
 \frac{2g_{\mu \nu} g_{\beta \gamma}-g_{\mu \beta} g_{\nu \gamma} -g_{\mu \gamma} g_{\nu \beta}} {(q_1 ^2-M^2)(q_2 ^2-M^2)} \nonumber
\end{eqnarray}

These four terms (half of $T_{24}$) summed with T1F,  the final result is
\begin{eqnarray}
\label{r1}
~&~& \int d^4q_1 d^4q_2 d^4q_3  (2 \pi)^4 \delta(P-q_1+q_2) \delta(q_1-k_1-q_3)\delta(q_3-k_2-q_2)\nonumber \\
~&\times & \frac{-6 k_1\cdot k_2 g_{\mu \nu} + 6 k_{2\mu}k_{1\nu} + 3M^2 g_{\mu \nu} + (-3 q_3^2 g_{\mu \nu}+12 q_{3\mu} q_{3\nu})} {(q_1 ^2-M^2)(q_3 ^2-M^2)(q_2 ^2-M^2)}
\end{eqnarray}
In D dimension, only difference is that the terms in the last bracket of the numerator of second line  is replaced by
$-(D-1) q_3^2 g_{\mu \nu}+4 (D-1) q_{3\mu} q_{3\nu}$.  The dealing with those of $T_3$ is just the same.


Similar way of calculation employing Feynman-Schwinger parametrization and Dyson subtraction as GWW    will give the exact same result as GWW. We will not
write in details here. And we will show how to get the gauge invariant result without the Dyson subtraction, considering the surface term of  the log divergence with the propagator inherent boundary condition at infinity in momentum space.

\section{tensor integral reduction,  surface term and physical boundary condition at infinity in momentum space}
Eq. (\ref{r1})  still has  logarithmic divergence to be canceled to get the finite result, 
  %
and simply summed with those from $T_3$ can not symmetrize {\it each} components of $q_3$.
  In  logarithmic divergent terms, the shift is allowed and one can just choose an independent loop momentum freely \cite{Jauch:1976ava}, 
   so one can use Feynman-Schwinger  parametrization to do the calculation,
as done by GWW.   
  In the GWW calculation, the na\"ive symmetrical integration in 4 dimension
  $$ \int d^4 l \frac{l^2 g_{\mu \nu}-4l_\mu l_\nu}{(l^2-M^2+i \epsilon)^3}=0$$
  is employed, similar as to calculate  T1LG.
  This is the key point which leads to  T1LG=0 and the requirement for Dyson subtraction in Eq. (\ref{r1}) to get the U(1) gauge invariant form by GWW.

 On the other hand,  for this logarithmic divergence,  one can first do the calculation in D dimension;  
   then 
   extrapolated  to 4 dimension, an extra term other than  doing na\"ive symmetrical integration in 4 dimension as above
      $$ \lim_{D \to 4} \int d^D l \frac{l^2 g_{\mu \nu}-4l_\mu l_\nu}{(l^2-M^2+i \epsilon)^3}=\frac{-i \pi^2}{2} g_{\mu \nu}$$
     can just play the role
 of Dyson subtraction term to let Eq. (\ref{r1}) invariant under U(1) gauge and get the similar result. However, this   suggests that TILG
 also must be calculated in D dimension and  
  extrapolated to 4 dimension. This  gives a finite terms,
 not zero  as the symmetrical integration, but
 just the non zero term at $ M \to  0$  \cite{egn,ikh,svvz}.  
 In the papers of GWW and  in other literatures  \cite{SVVZ12, HTW, MZW,Jegerlehner:2011jm, SZC, CCNS, pit}, all the authors agree with these mathematical results,
 and clearly pointed out that the dealing with the logarithmic divergence
 in 4 dimension without regularization by GWW or dealing with a regularization scheme (e.g., dimensional regularization) by others is
   the reason of the contradiction on the non-zero term for $ M \to 0$
 for this specific process in the unitary gauge.


To understand whether or not adopt the GWW na\"ive symmetrical integration in 4 dimension,
here only the following key point has to be discussed,
which is pointed out long ago but not fully analyzed:  
 The logarithmically  integration (in the following we omit the $+i\epsilon$ term since we  explicitly discuss the integration contour)
\begin{equation}
\int d^4 l \frac{l^\mu l^\nu}{(l^2-\Delta)^3},
\end{equation} is related with the surface term by
 %
\begin{equation}
\partial_l^\mu \frac{l^\nu}{(l^2-\Delta)^2}=\frac{g^{\mu \nu}}{(l^2-\Delta)^2}-\frac{4 l^\mu l^\nu}{(l^2-\Delta)^3}=\frac{l^2 g^{\mu \nu}- 4 l^\mu l^\nu}{(l^2-\Delta)^3}-\frac{\Delta g^{\mu \nu}}{(l^2-\Delta)^3}.
\end{equation}
So, we have
\begin{equation}
\int d^4 l \frac{l^2 g^{\mu \nu}- 4 l^\mu l^\nu}{(l^2-\Delta)^3}= \int d^4 l \partial_l^\mu \frac{l^\nu}{(l^2-\Delta)^2}+ \int d^4 l \frac{\Delta g^{\mu \nu}}{(l^2-\Delta)^3}.
\end{equation}
In the  above equation, employing the Gaussian theorem in 4-dimension space-time, the first term at the  right side is
\begin{equation}
\label{surf}
\int d^4 l \partial_l^\mu \frac{l^\nu}{(l^2-\Delta)^2} = \int d\sigma^\mu \frac{l^\nu}{(l^2-\Delta)^2}.
\end{equation}
As the Gaussian theorem allow, the 3-dimension hypersurface boundary can be as far away as possible.
This integral equals to zero, which is a nontrivial conclusion by the definition  of free Feynman propagator.

As we all know, the Feynman  propagator is defined as a contour integral of the complex $k^0$ plane. To have a definite result, consistently include the physical poles,  one must require to include the  $k^0 >|\vec{k}|$ region. This means that the boundary for the 4 dimension region must lie in the place with $k^0 >|\vec{k}|$.    For employing the Gaussian theorem and to be consistent
with this free Feynman propagator, one can even further require a boundary in infinity  with the $k^0$ much much larger than $|\vec{k}|$.  The reason is that the  hyper surface of the Gaussian theorem can be much much more larger than 'necessary', only requirement is that it covers the integration volume.  So in such a hypersurface the integrand is vanishing and we get the zero surface term. This also precisely define the boundary condition of the Dyson scheme.  We can put the boundary hypersurface like this because the 'elements' in  perturbative calculation in Dyson scheme of interaction picture  is the free theory (propagator/Green function), free spacetime, free vacuum, and scattering theory is a static solution. This solution for a single particle is described as summation of plane wave and sphere wave, as the  solution of the homogeneous and inhomogeneous Helmholtz Equations, respectively.  In this standard case we can put the k as large as possible, also put $k^0$ as large as possible, and require $k^0$ much larger than k. Hence the 'most far', especially most far for $k^0$ hypersurface integral as Eq. (\ref{surf})  to be zero.

By the above discussion, we see in 4-dimension Minkowski space-time,
\begin{equation}
\int d^4 l \frac{l^2 g^{\mu \nu}- 4 l^\mu l^\nu}{(l^2-\Delta)^3}=  \int d^4 l \frac{\Delta g^{\mu \nu}}{(l^2-\Delta)^3}=- i \frac{\pi ^2}{2} g^{\mu \nu}
\end{equation}
Or,
\begin{equation}
\int d^4 l \frac{l^\mu l^\nu}{(l^2-\Delta)^3}= \frac{1}{4} \int d^4 l \frac{l^2 g^{\mu \nu}}{(l^2-\Delta)^3}+ i \frac{\pi ^2}{8} g^{\mu \nu}
\end{equation}

\begin{equation}
\int d^4 l \frac{l^\mu l^\nu}{(l^2-\Delta)^3}= \frac{1}{4} \int d^4 l \frac{ g^{\mu \nu}}{(l^2-\Delta)^2}
\end{equation}
This result gives an explicit example of the difference of Minkowski spacetime from the Euclid one. For the latter,  one may take the most simple periodic boundary condition with the time component $k^0$ at the same footing as the space components, and hence a straightforward replacement $l^\mu l^\nu \to \frac{1}{4} l^2 \delta^{\mu \nu}$ can be taken. But this can not keep the relation $k^0 > |\vec{k}|$, which  must be respected since our standard scattering theory of the Dyson scheme is a four dimension covariant (Minkowsi) realization of the stationary solution of the Helmholtz Equation.

 Here we would like to point out that, however, for {\it finite} integral, the above interface term (=0) analysis is consistent with the simple replacement $l^\mu l^\nu \to \frac{1}{4} l^2 g^{\mu \nu}$. Dimensional regularization is self-consistent in this aspect, i.e., the replacement $l^\mu l^\nu \to \frac{1}{D} l^2 g^{\mu \nu}$ always equals to take the surface term zero for any integral, since in D dimension all integrals considered finite.  

Besides that gauge (both $SU(2)\times U_Y(1)$ and $U_{em}(1)$) invariant  finite result of $H \to \gamma \gamma$  can be obtained with calculation in 4 dimension Minkowski phase space,  without the introduction of the 'Dyson subtraction' and without regularization,   taking into account the above analysis of surface term, 
we have employed  the 
 above analysis to calculate the  $H \to \gamma Z$ process \cite{LW} as another example to demonstrate the  general way of Dyson scheme to set the loop momenta without the choice of the specific independent ones  and eschew shift.  There to deal with the quadratic divergence cancellation for terms proportional to
   $M_Z^2/M^4$,
we have to use the relation for the quadratic one
\begin{equation}
\label{qud}
\partial_{l}^\mu \frac{l^\nu}{(l^2-\Delta)}=\frac{g^{\mu \nu}}{(l^2-\Delta)}-\frac{2 l^\mu l^\nu}{(l^2-\Delta)^2}=\frac{l^2 g^{\mu \nu}- 2 l^\mu l^\nu}{(l^2-\Delta)^2}-\frac{\Delta g^{\mu \nu}}{(l^2-\Delta)^2}.
\end{equation}
By the help of  both the quadratic and log relations as above, we obtain the decay amplitude of $ H \to \gamma Z $ process
via one W loop in the unitary gauge.           %
      The divergent integrals including  those of high divergence orders typical of unitary gauge 
      are all  arranged to cancel to get the  electromagnetic $U(1)$ gauge invariant  finite result, hence no contribution to the  renormalization  constant of $Z\gamma$-mixing in this 1-loop subprocess.
    For the calculation of the Feynman diagrams employing the   Feynman rules, we can insist that all the integrations of   the  propagator momenta and  all the $\delta$-functions representing the 4-momentum conservation of every vertex
 are retained    in the beginning. Therefore,  the  ambiguity of  setting independent  loop momentum for  divergences worse than logarithmic does not exist, and shift of integrated variable in such  divergent integrals   is eschewed.  The calculation  also is insisted to be done in 4-dimension Minkowski momentum space without the aid of any regularization.
 The obtained form of the amplitude is like that of  $ H \to \gamma \gamma $ (both proportional to the U(1) invariant form $k_1\cdot k_2 g_{\mu\nu}-k_{2\mu}k_{1\nu}$), except extra  terms proportional to
   $M_Z^2/M^4$ (or say, $\frac{1}{M^2} \frac{1}{cos^2\theta_W}$) remained  after the cancellation of all divergences from quadratic to log. This part of terms  is crucial to study the non-Abelian gauge invariance for the $ H \to \gamma Z $ process.
   The correct treatment on the  surface terms for the quadratic and logarithmic tensor integral  is one of  the key points, especially in obtaining the above mentioned  finite $M_Z^2/M^4$ terms. Another key application of the correct surface term is to show that the Higgs ZZ direct coupling  with bubble and tadpole loop to $\gamma Z$ contributes zero.
   Furthermore in the calculations subtle cancellations  between terms from different origins  not only show the self-consistency of standard model but also the unitary gauge, besides the Dyson scheme. The unitary gauge can then be used to do more complex loop calculations. Less diagrams and clear physical picture are the advantages.

The  {\it Feynman} propagator, is the vacuum expectation value of the time ordered product of field operators.
The time-boundary condition for these   operators are as of the free field since the interaction is adiabatically removed at time  to plus/minus infinity, and
for interactions decreasing fast enough with distance, the wave function in infinitely distant place should be the  superposition of the plane wave and the spherical  wave.
Because of this, first of all,  in the 4-dimension expression in momentum space, two poles of the denominator guarantee
 the Heaviside function for the time order, hence for the proper causality for the framework of perturbative theory   in the interaction picture of the canonical quantization. So the integral region of momentum space, 
 especially for large  values, must guarantee to accommodate the proper contour to take the physical poles.

 Let' s see this topic in more details. The coordinate space Feynman propagator (takeing a scalar field as example) is
 \begin{equation}
 \label{pro}
 D_F(x-y)=\int \frac{d^4k}{(2 \pi)^4} \frac{i}{k^2-m^2+i \epsilon} e^{-ik\cdot (x-y)}.
 \end{equation}
 In doing the integration of $k^0$ on the complex number plane,
the    behaviour of the integrand when $k^0$ goes to infinity  must be suppressed,  by the imagine part of $k^0$, via the (negative real) exponential factor.
This is convinced by the Jordan Lemma. Indeed, the exponential factor in the integral of Eq. (\ref{pro}) is the necessary factor for the application of the residue theorem to do the integration of  Eq. (\ref{pro}).
Hence for a specific Feynman diagram, though the behaviour of the integrand  in the phase space could be  complex, but for  each  $q$, the momentum of the propagator,
the large $q^0$ behaviour must be suppressed, or else the definition of the propagator is deviated from what is in the original form, such as  Eq. (\ref{pro}). At the same time, for the pole
to be in the closed contour so as to get non zero result, we must require the modulo:
\begin{equation}
\lim_{q^0 \rightarrow \infty, q \rightarrow \infty} ~~~\frac{q}{q^0} < 1.
\end{equation}
So the behaviour of the $q^0$ control all the four components in infinity.  
And this means, if without the exponential functions, or once it is integrated to get  the $\delta$ functions, one has to keep in mind
  the interaction picture perturbative theory in fact based on the on shell real physical (free) particle Hilbert space. The 'virtual off-shell states'
  mainly come from the way to write the propagator from a three dimensional  integral to the four dimensional covariant one.
 The extra integration is introduced as the contour integral on the complex plane. With the condition in the above equation to keep the pole in the contour,
   the infinite behaviour is 'enforced' to be suppressed  by the exponential, so that the Jordan lemma valid and the 3-dimensional and  4-dimensional  forms equal to each other.
 This indicates that,    
 without the exponential,  the  behaviour at infinity of the propagator momenta,  is still in need to be suppressed, to fulfill the requirement of the Jordan Lemma.
   %
   For worse divergent terms
 that exactly the same and cancel, we need not to take account this fact.
  But this fact is needed as the basis for  doing  symmetrical integration,  convergent or divergent, to obtain the correct surface term.
%
%
In the realistic  calculation, the Wick rotation is performed. Before the rotation, since the exponential factor, as convinced by the  Jordan lemma,
integration of $k^0 $ on the infinite points  of the imaginary axis is suppressed and this is also the necessary condition for the rotation, i.e., no poles in 1,3 quadrants of the complex $k^0$ plane. After the rotation, these four components are in more symmetric footing. The 'periodic (or  vanishing) boundary conditions'  for space components is similar,   as to considering the Lippmann-Schwinger Equation for the static scattering state.
The three dimensional Green function of the three dimensional Lippmann-Schwinger Equation is also defined by a contour integral  on the complex momentum plane  with a exponential, to guarantee
the wave function in infinite distance is the superposition of the plane wave and the spherical wave, when the interaction potential decrease fast enough.
The exponential suppresses the infinite  contribution of the momentum. This should be kept with or without the Wick rotation.

The Dyson scheme, the way re-introduced in this paper, especially not to integrate the $\delta$ functions before have to,
provides the exact definition of the Feynman diagrams free from the uncertainty of momentum shift and/or of  various ways of setting independent integrated loop momenta. 
It  in fact provides the necessary way to let  the unphysical ultraviolet  divergences cancelled, even in case that
the inter-relations among various Feynman diagrams which has investigated long ago \cite{MY, TTW} and reconsidered in GWW  may not be applicable.
Though in general regularization is  very important practical tool, and in general redefining QFT or its perturbative  scattering  (static) theory in various ways can explore new developments, for the concrete process $H \to \gamma \gamma$ via one W loop in standard model,
 the original  Dyson scheme can definitely assign the loop momenta, and give the SU(2) $\times$ U(1) gauge invariant result without the need of any regularization  or  redefinition of QFT.
 To get the result, a specific boundary condition in momentum space inherent in the Dyson scheme is exposed, which makes sense in various problems, e.g., the conservation of the axial vector current \cite{Bao:2021byx}.
All these works, including that of $\gamma Z$, indicate that the  Dyson S-matrix theory is much more self-consistent than expected before.
The finite 'uncertainty' of the  logarithmic  and quadratic divergent integrals are related with the boundary condition of the infinite momentum surface. This straightforwardly indicates that the singularities of the spacetime will make sense, especially when the spacetime is of  complex connection in the early time of the universe, which can lead the boundary condition {\it beyond the Dyson case}.  For example, this can results in a large axial vector current  anomaly which can give large strong interaction CP violation within standard model, attractive for investigating baryogenesis. 
That is to say, the above discussion on the (ultraviolet) divergence cancellation possibly depending on the concrete (local) structure of spacetime implies interesting speculation.  The 'historical versions'  of  QFT's vary with the  universe  evolution. QFT's  can be defined on various spacetime math structure. In more details,
 different versions of QFT's, defined on different manifolds or more general math structures corresponding to our universe---spacetime---at various special periods can be self -consistent or -inconsistent. 
 But in whatsoever cases,
the divergences may cancel to get finite (including zero) prediction or may not and can not give clear prediction, it can always  probe the math structure of the spacetime of the universe. The ultraviolet probes the local, the infrared probes the global, or just manifests the special information of the space time which leads to the inconsistency between the spacetime and QFT.
In the more concrete cases addressed above in the paper, the spacetime in early universe manifold may have singularity, defect, bubble wall or other structure to cause the anomaly to give the CP violation for baryogenesis but these structures of the spacetime manifold now evolute/expand so that locally become Minkowski and no anomaly, no violations but only conservation of the corresponding currents, which is consistent with experiments and has to ask for cancellation in some theoretical  framework without  elimination inherently.  At the same time these early defects or structures can also play the role of  'seed' of the curvature of space time via the homogeneous Einstein equations, i.e., this curvature is  not caused by 'matter' (inhomogeneous term in Einstein equations) but is now observed as 'dark matter' when the universe evolute to today.

\acknowledgments
The author deeply thanks  Prof. Tai Tsun Wu for stimulating these topics and  for numerous instructive discussions.    The hospitality of CERN TH unit is also greatly thanked for providing short-term visitor opportunities.
This work is supported in part by  National Natural Science Foundation of China (grant No. 12275157, 11775130, 11635009).



\appendix
 \section{Mathematics in   term of the   $H \to \gamma Z$ process corresponding to Eqs. (2.5-2.12) of GWW \cite{Gastmans:2011wh}}
(These formulation  will recover to the complemented  GWW formulations which we employed in the paper for  calculating the $H \to \gamma \gamma $ process once taking $M_Z=0$.)
\begin{equation}
\label{a1}
k_1^2=0, ~ k_2^2=M_Z^2; \hspace{ 3 cm}  k_{1\mu}=k_{2\nu}=0.
\end{equation}
\begin{equation}
(k_1+k_2)^2=2 k_1 \cdot k_2 +M_Z^2=M_H^2.
\end{equation}
\begin{eqnarray}
V_{\alpha \beta \gamma} (p_1, p_2, p_3)= (p_2-p_3)_{\alpha}g_{\beta \gamma} +(p_3-p_1)_{\beta}g_{\gamma \alpha}+(p_1-p_2)_{\gamma}g_{\alpha \beta}; \\
p_1+p_2+p_3=0~~ (incoming). \nonumber
\end{eqnarray}
\begin{eqnarray}
p_1^{\alpha} V_{\alpha \beta \gamma} (p_1, p_2, p_3)=(p_3^2g_{\beta \gamma}-p_{3 \beta}p_{3 \gamma})-(p_2^2g_{\beta \gamma}-p_{2 \beta}p_{2 \gamma})\\
V_{\alpha \beta \gamma} (p_1, p_2, p_3) p_3 ^\gamma =-(p_1^2 g_{\alpha \beta}-p_{1\alpha} p_{1\beta}) +(p_2^2 g_{\alpha \beta}-p_{2\alpha} p_{2\beta}) \nonumber
\end{eqnarray}
\begin{eqnarray}
p_1^{\alpha} V_{\alpha \mu \gamma} (p_1, -k_1, p_3)=p_3^2g_{\mu \gamma}-p_{3 \mu}p_{3 \gamma}\\
V_{\alpha \mu \gamma} (p_1, -k_1, p_3) p_3 ^\gamma =-(p_1^2 g_{\alpha \mu}-p_{1\alpha} p_{1\mu})\nonumber \\
p_1^{\alpha} V_{\alpha \nu \gamma} (p_1, -k_2, p_3)=p_3^2g_{\nu \gamma}-p_{3 \nu}p_{3 \gamma}-M_Z^2 g_{\nu \gamma}\\
V_{\alpha \nu \gamma} (p_1, -k_2, p_3) p_3 ^\gamma =-(p_1^2 g_{\alpha \nu}-p_{1\alpha} p_{1\nu})+M_Z^2 g_{\alpha \nu}\nonumber
\end{eqnarray}
\begin{eqnarray}
\label{a7}
p_1^{\alpha} V_{\alpha \mu \gamma} (p_1, -k_1, p_3)=(p_3^2-M^2)g_{\mu \gamma}-p_{3 \mu}p_{3 \gamma}+M^2 g_{ \mu \gamma}\\
V_{\alpha \mu \gamma} (p_1, -k_1, p_3) p_3 ^\gamma =-[(p_1^2-M^2) g_{\alpha \mu}-p_{1\alpha} p_{1\mu}]-M^2 g_{\alpha \mu }\nonumber \\
p_1^{\alpha} V_{\alpha \nu \gamma} (p_1, -k_2, p_3)=(p_3^2-M^2)g_{\nu \gamma}-p_{3 \nu}p_{3 \gamma}+(M^2-M_Z^2) g_{ \nu \gamma}\\
V_{\alpha \nu \gamma} (p_1, -k_2, p_3) p_3 ^\gamma =-[(p_1^2-M^2) g_{\alpha \nu}-p_{1\alpha} p_{1\nu}]-(M^2-M_Z^2) g_{\alpha \nu }\nonumber
\end{eqnarray}
\begin{equation}
p_1^{\alpha} V_{\alpha \mu \gamma} (p_1, -k_1, p_3) p_3 ^\gamma=0
\end{equation}
\begin{equation}
\label{a10}
p_1^{\alpha} V_{\alpha \nu \gamma} (p_1, -k_2, p_3) p_3 ^\gamma=-M_Z^2 p_{3\nu}=M_Z^2p_{1\nu}
\end{equation}
 %
\section{$R_1$ gauge}
 %
 %
The goldstone contribution (all the W propagators are replaced by the charged goldstone ghost propagators in FIG. 1) is the only ones which can contribute to the term not zero  for $M=0$, they are
(times the $-ie^2g M_H^2/ M$  factor):   
 \begin{equation}
 A=  \int d^4q_1 d^4q_2 d^4q_3 \frac{  (2 \pi)^4 \delta(P-q_1+q_2) \delta(q_1-k_1-q_3)\delta(q_3-k_2-q_2) }  {(q_1 ^2-M^2)(q_3 ^2-M^2)(q_2 ^2-M^2) }(4 q_{3\mu} q_{3 \nu})
 \end{equation}
  \begin{equation}
 B=  \int d^4q_1 d^4q_2 d^4q_3 \frac{  (2 \pi)^4 \delta(P-q_1+q_2) \delta(q_1-k_1-q_3)\delta(q_3-k_2-q_2) }  {(q_1 ^2-M^2)(q_3 ^2-M^2)(q_2 ^2-M^2) }(-q_3^2g_{\mu \nu})
 \end{equation}
 \begin{equation}
 C=  \int d^4q_1 d^4q_2 d^4q_3 \frac{  (2 \pi)^4 \delta(P-q_1+q_2) \delta(q_1-k_1-q_3)\delta(q_3-k_2-q_2) }  {(q_1 ^2-M^2)(q_3 ^2-M^2)(q_2 ^2-M^2) }(M^2g_{\mu \nu})
 \end{equation}
 After the Feynman-Schwinger parametrization,  A contributes $a1=4 l_{\mu} l_{\nu}$ and  $a2=-4 \alpha_1 \alpha_2 k_2 \mu k_1 \nu$.  B contributes $b1=- l^2 g_{\mu \nu}$ and  $b2=2 \alpha_1 \alpha_2 k_1 \cdot k_2 g_{\mu \nu}$. $\alpha_1 \alpha_2$ are the Feynman-Schwinger parameters.
  C contributes  $M^2 g_{\mu \nu}$.



  Only the convergent term of A and B, $a2$ $b2$  can not give the U(1) gauge invariant amplitude. However, both
   the dimensional regularization calculation 
   and the 4 dimension calculation with the correct surface term  on the log divergent
   $ (a1+b1)$  can compensate  to give the U(1) gauge invariant result. On the other hand, to take the na\"{i}ve symmetrical integration in 4 dimension, e.g., the case of  na\"{i}ve cutoff,  where log $ (a1+b1) \to 0$, then   the Dyson subtraction term (take the $M_H$ as coupling constant not zero) can help to   give the same result. I.e.,  All three cases
   give
  \begin{equation}
  g_{\mu \nu}(\frac{M^2}{M^2-2 \alpha_1 \alpha_2k_1\cdot k_2}-1)=  g_{\mu \nu}\frac{2\alpha_1 \alpha_2k_1\cdot k_2 }{(M^2-2 \alpha_1 \alpha_2k_1\cdot k_2)}
  \end{equation}
  and can combine with the convergent terms  to recover the U(1) gauge invariance and give the non zero term for M=0.

  So, in the unitary gauge,  employing dimensional regularization or   na\"{i}ve symmetrical integration with the  Dyson subtraction will  give different results, as shown in above and by GWW. In
    $R_1$ gauge, these two approaches  give the similar result. 
  %
However, the Dyson scheme including the correct surface term will always give the definite result, for this specific problem consistent with dimensional regularization  and  (Abelian \& non-Abelian) gauge invariant.
So the 'Dyson subtraction' is not always possible to work as the necessary or sufficient compensation. As a matter of fact, the Dyson subtraction was introduced in his classical paper to deal with the photon photon scattering
fermion-loop box diagram to get the  gauge invariant result. But that case is also a problem of high order divergence, and should  better be calculated with the Dyson scheme to eliminate the loop momentum uncertainty. Here we would like to remind that the divergence of a loop diagram include
every ways to go to infinity by the loop momenta. Therefore the divergence can be higher than the superficial one by 'power counting'.  One should notice that, e.g.,  the N ($>$3) point one fermion loop can always be linearly divergent, no matter how large N. The reason lies in the structure of the fermion propagator.
By taking a light-cone form of  4-momentum, one can easily find the momentum region this loop diagram goes to infinity linearly \cite{chw}. Regularization sometimes sets a special way for the loop momenta going to infinity, which has to be
discussed in a most general consideration.


\section{Standard Ward identity in QED}
 From the above we learn that
for any diagram, there is a definitely 'original' definition  on  loop momenta in  the  integral without any uncertainty. This experience of calculation
demonstrated in this paper is expected to be applied to multi-loop diagram calculations to eliminate any ambiguity  for the choice of the  integrated loop momenta.

Besides, this also helps for  understanding  the momentum flow in  the Ward identity of QED \cite{MY}.

Things can be easily made clear at one loop for QED.  The electron self energy can be written as
 \begin{equation}
 \label{sig}
 \Sigma(p_1)=\Sigma(p_1, p_2) \sim \int d^4q d^4 k \delta(p_1-q-k) \delta(q+k-p_2) \gamma^\mu \frac{1}{\cancel{q}-m} \gamma_\mu \frac{1}{k^2}
 \end{equation}
It represent a factor like that in the square bracket of the following equation
\begin{equation}
\int dx \int dy \cdot \cdot \cdot ~~  \cdot \cdot \cdot \bar{\Psi}(x) [ A^\cdot _\mu(x)\gamma^\mu \Psi^\P(x) \bar{\Psi}^\P (y)A^\cdot_\nu(y)\gamma^\nu ] \Psi (y)  \cdot \cdot \cdot
\end{equation}
as a part of  the Wick expansion of some order of the perturbation expansion of the S-matrix (the $^\cdot$ and $^\P$ signalize the 'contraction', i.e., vacuum expectation value of the time-ordered product of the field operators).
 We can move  one $\delta$ function $\delta(q+k-p_2)=\delta(p_1-p_2)$ out, 
  only deal with
 \begin{equation}
 \Sigma(p_1) \sim \int d^4q d^4 k \delta(p_1-q-k)  \gamma^\mu \frac{1}{\cancel{q}-m} \gamma_\mu \frac{1}{k^2},
 \end{equation}
and in deed track back to the original form
 \begin{equation}
 \Sigma(p_1)\sim \int d^4 x_1 \int d^4q d^4 k e^{ i(p_1-q-k) \cdot x_1}  \gamma^\mu \frac{1}{\cancel{q}-m} \gamma_\mu \frac{1}{k^2}
 \end{equation}
 (($2\pi)^4$ factor taken away).

 So we have
  \begin{eqnarray}
 \frac{\partial \Sigma(p_1)}{\partial p_{1\nu}} &\sim& \int d^4 x_1 \int d^4q d^4 k ~(i x_1 ^\nu) e^{ i(p_1-q-k) \cdot x_1}  \gamma^\mu \frac{1}{\cancel{q}-m} \gamma_\mu \frac{1}{k^2}\nonumber \\
~ &=& \int d^4 x_1 \int d^4q d^4 k ~(- \frac{\partial}{\partial q_\nu} e^{ i(p_1-q-k) \cdot x_1})  \gamma^\mu \frac{1}{\cancel{q}-m} \gamma_\mu \frac{1}{k^2}\nonumber \\
~ &=& \int d^4 x_1 \int d^4q d^4 k  e^{i(p_1-q-k) \cdot x_1}  \gamma^\mu (\frac{\partial}{\partial q_\nu} \frac{1}{\cancel{q}-m}) \gamma_\mu \frac{1}{k^2}
 \end{eqnarray}
with a surface term eliminated for getting the last line (This is consistent with the static single particle wave function space boundary condition referring to Lippman-Schwinger Equation).
This explains why seems $p_1$ only flow via the fermion line, hence the partial derivative can only operate on the fermion lines (of course it can only flow via the photon line, to get a 'useless'  formulation not relevant to the W.I.) 
not both the photon and fermion lines, the latter may lead to a double counting.
This conclusion can be simply read from Eq. (\ref{sig}) formally as
\begin{eqnarray}
&~&\frac{\partial \Sigma(p_1,p_2)}{\partial p_{1\nu}}=\frac{\partial \Sigma(p_1,p_2)}{\partial p_{2\nu}}, ~ and ~ since~  p_1=k+q=p_2~ before~ integration, \\
&=&\int d^4q d^4 k \frac{\partial }{\partial (q+k)_{\nu}} [ \delta(p_1-q-k) \delta(q+k-p_2) \gamma^\mu \frac{1}{\cancel{q}-m} \gamma_\mu \frac{1}{k^2}]\\
&=&\int d^4q d^4 k \frac{\partial }{\partial q_{\nu}} [ \delta(p_1-q-k) \delta(q+k-p_2) \gamma^\mu \frac{1}{\cancel{q}-m} \gamma_\mu \frac{1}{k^2}]\\
&=&\int d^4q d^4 k \frac{\partial }{\partial k_{\nu}} [ \delta(p_1-q-k) \delta(q+k-p_2) \gamma^\mu \frac{1}{\cancel{q}-m} \gamma_\mu \frac{1}{k^2}]\\
& \neq & \int d^4q d^4 k (\frac{\partial }{\partial q_{\nu}} +\frac{\partial }{\partial k_{\nu}})[ \delta(p_1-q-k) \delta(q+k-p_2) \gamma^\mu \frac{1}{\cancel{q}-m} \gamma_\mu \frac{1}{k^2}].
\end{eqnarray}
Since the starting point expression of such one loop self energy is linear divergent, based on our discussions in the  previous sections, deliberately setting different
independent integrated loop momentum will lead to ambiguity. It is the consistency with respect to  the Ward Identity determines the general choice, i.e.,
using $k$ in the above integration as the   independent integrated loop momentum, and taking $q=p_1-k$, $p_1$ only flowing through the fermion line.




\begin{thebibliography}{150}


\bibitem{wein73}
S. Weinberg, Phys. Rev. D 7 (1973) 1068.

\bibitem{Wu:2017rxt}
  T.~T.~Wu and S.~L.~Wu,
  Nucl.\ Phys.\ B {\bf 914} (2017) 421.
  doi:10.1016/j.nuclphysb.2016.11.007

\bibitem{Wu:2016nqf}
  T.~T.~Wu and S.~L.~Wu,
  Int.\ J.\ Mod.\ Phys.\ A {\bf 31} (2016) no.04n05,  1650028.
  doi:10.1142/S0217751X16500287

\bibitem{Gastmans:2015vyh}
  R.~Gastmans, S.~L.~Wu and T.~T.~Wu,
  Int.\ J.\ Mod.\ Phys.\ A {\bf 30} (2015) no.32,  1550200.
  doi:10.1142/S0217751X15502000

\bibitem{Gastmans:2011wh}
  R.~Gastmans, S.~L.~Wu and T.~T.~Wu,
  arXiv:1108.5872 [hep-ph].

\bibitem{Gastmans:2011ks}
  R.~Gastmans, S.~L.~Wu and T.~T.~Wu,
  arXiv:1108.5322 [hep-ph].


\bibitem{Dyson:1949ha}
  F.~J.~Dyson,
  Phys.\ Rev.\  {\bf 75} (1949) 1736.
  doi:10.1103/PhysRev.75.1736



\bibitem{SVVZ12}
M.~Shifman, A.~Vainshtein, M.~B.~Voloshin and V.~Zakharov,
  Phys.\ Rev.\ D {\bf 85}, 013015 (2012).

\bibitem{HTW}
D.~Huang, Y.~Tang and Y.~L.~Wu,
  Commun.\ Theor.\ Phys.\  {\bf 57}, 427 (2012).

\bibitem{MZW}
W.~J.~Marciano, C.~Zhang and S.~Willenbrock,
  Phys.\ Rev.\ D {\bf 85}, 013002 (2012).

\bibitem{Jegerlehner:2011jm}
  F.~Jegerlehner,
  Comment on $H \to \gamma \gamma$ and the Role of the Decoupling Theorem and the Equivalence Theorem,
  arXiv:1110.0869 [hep-ph].

\bibitem{SZC}
H.~S.~Shao, Y.~J.~Zhang and K.~T.~Chao,
  JHEP {\bf 1201}, 053 (2012).

\bibitem{CCNS}
A.~L.~Cherchiglia, L.~A.~Cabral, M.~C.~Nemes and M.~Sampaio,
Phys.\ Rev.\ D {\bf 87}, no.\ 6, 065011 (2013).

\bibitem{pit}
A.~M.~Donati and R.~Pittau,
  JHEP {\bf 1304}, 167 (2013).


\bibitem{Melnikov:2016nvo}
  K.~Melnikov and A.~Vainshtein,
  Phys.\ Rev.\ D {\bf 93} (2016) no.5,  053015.

\bibitem{Christova:2014mea}
  E.~Christova and I.~Todorov,
  Bulg.\ J.\ Phys.\  {\bf 42}, 296 (2015).



\bibitem{egn}
J.~R.~Ellis, M.~K.~Gaillard and D.~V.~Nanopoulos,
  Nucl.\ Phys.\ B {\bf 106}, 292 (1976).

\bibitem{ikh}
B.L.~Ioffe and V.A.~Khoze,
  Sov.\ J.\ Part.\ Nucl.\  {\bf 9}, 50 (1978)
  [Fiz.\ Elem.\ Chast.\ Atom.\ Yadra {\bf 9}, 118 (1978)].

\bibitem{svvz}
M.A.~Shifman, A.I.~Vainshtein, M.B.~Voloshin and V.I.~Zakharov,
  Sov.\ J.\ Nucl.\ Phys.\  {\bf 30}, 711 (1979)
  [Yad.\ Fiz.\  {\bf 30}, 1368 (1979)].

\bibitem{MY}
R. L. Mills and C. N. Yang, Prog. Theo. Phys. Suppl. 37 (1966) 507.

\bibitem{TTW}
T. T. Wu, Phys. Rev. 125 (1962) 1436.

\bibitem{plbat}
ATLAS Collab., Phys. Lett. B 732 (2014) 8-27.

\bibitem{plbcm}
CMS Collab.,  Phys. Lett. B 726 (2013) 587-609.


\bibitem{LW}
 'Amplitude of $H \to \gamma Z$ process via one W loop in unitary gauge',  to be submitted by the author.

\bibitem{Bao:2021byx}
S.~S.~Bao, S.~Y.~Li and Z.~G.~Si,
[arXiv:2109.14835 [hep-ph]].

\bibitem{Ferreira:2011cv}
L.~C.~Ferreira, A.~L.~Cherchiglia, B.~Hiller, M.~Sampaio and M.~C.~Nemes,
Phys. Rev. D \textbf{86} (2012), 025016
doi:10.1103/PhysRevD.86.025016
[arXiv:1110.6186 [hep-th]].


\bibitem{Duch:2020was}
P.~Duch, M.~D\"utsch and J.~M.~Gracia-Bond\'\i{}a,
Eur. Phys. J. C \textbf{81} (2021) no.2, 131
doi:10.1140/epjc/s10052-021-08898-z
[arXiv:2011.12675 [hep-ph]].



\bibitem{Jauch:1976ava}
J.~M.~Jauch and F.~Rohrlich,
Springer, 1976,
ISBN 978-3-642-80953-8, 978-3-642-80951-4
doi:10.1007/978-3-642-80951-4

\bibitem{chw}  H. Cheng and T. T. Wu, Expanding Protons: Scattering at High Energies, MIT Press, Cambridge, MA, 1987.

\end{thebibliography}
\end{document}